\documentclass[letter]{article}
\pdfoutput=1 

\usepackage{jheppub} 
\usepackage{hyperref}
\usepackage[T1]{fontenc} 
\usepackage[normalem]{ulem}
\usepackage{graphicx}
\usepackage{caption}
\usepackage{subcaption}
\usepackage{dcolumn}
\usepackage{bm}
\usepackage{amsmath}
\usepackage{amssymb}
\usepackage{multirow}
\usepackage{color,xcolor,url,ulem,verbatim,appendix}
\usepackage[colorlinks=true
,urlcolor=blue
,anchorcolor=blue
,citecolor=blue 
,filecolor=blue
,linkcolor=blue
,menucolor=blue
,linktocpage=true
,pdfproducer=medialab
,pdfa=true
]{hyperref}
\usepackage{slashed}
\newcommand{\beq}{\begin{equation}}
\newcommand{\eeq}{\end{equation}}
\newcommand{\bea}{\begin{eqnarray}}
\newcommand{\eea}{\end{eqnarray}}
\newcommand{\no}{\nonumber}

\pdfoutput=1 

\title{\begin{center}
\LARGE{\textbf{Conformal Leptogenesis in Composite Higgs Models}}

\end{center}
}
\author[a]{Kaustubh Agashe}
\emailAdd{kagashe@umd.edu}
\author[b]{Peizhi Du}
\emailAdd{peizhi.du@rutgers.edu}
\author[c]{Majid Ekhterachian}
\emailAdd{majid.ekhterachian@epfl.ch}
\author[d]{Chee Sheng Fong}
\emailAdd{sheng.fong@ufabc.edu.br}
\author[e]{Sungwoo Hong}
\emailAdd{sungwooh@kaist.ac.kr}
\author[f]{Luca Vecchi}
\emailAdd{luca.vecchi@pd.infn.it}
\affiliation[a]{Maryland Center for Fundamental Physics, Department of Physics,\\ University of Maryland, College Park, MD 20742 USA}
\affiliation[b]{New High Energy Theory Center, Department of Physics and Astronomy, Rutgers University, Piscataway, NJ 08854, USA}
\affiliation[c]{Theoretical Particle Physics Laboratory (LPTP),
Institute of Physics, EPFL, Lausanne, Switzerland}
\affiliation[d]{Centro de Ciências Naturais e Humanas,
Universidade Federal do ABC, 09.210-170,\\ Santo André, SP, Brazil}
\affiliation[e]{Department of Physics, Korea Advanced Institute of Science and Technology (KAIST), Daejeon 34141, Republic of Korea}
\affiliation[f]{Istituto Nazionale di Fisica Nucleare (INFN), Sezione di Padova, Italy}

\abstract{
We study the generation of the baryon asymmetry in Composite Higgs  models with partial compositeness of the Standard Model (SM) fermions and heavy right-handed neutrinos, developing for the first time a complete picture of leptogenesis in that setup. 
The asymmetry is induced by the out of equilibrium decays of the heavy right-handed neutrinos into a plasma of the nearly conformal field theory (CFT), i.e. the deconfined phase of the Composite Higgs dynamics. This exotic mechanism, which we call \emph{Conformal Leptogenesis}, admits a reliable description in terms of a set of  ``Boltzmann equations'' whose coefficients can be expressed in terms of correlation functions of the CFT. The asymmetry thus generated is subsequently affected by the supercooling resulting from the confining phase transition of the strong Higgs sector as well as by the washout induced by the resonances formed after the transition. Nevertheless, a qualitative description of the latter effects suggests that conformal leptogenesis can successfully reproduce the observed baryon asymmetry in a wide region of parameter space. A distinctive signature of our scenarios is a sizable compositeness for {\em all} the generations of SM neutrinos, 
which is currently consistent with all constraints but may be within reach of future colliders.
}
\preprint{UMD-PP-024-10}

\begin{document}
\hspace{22em} 
\maketitle
\flushbottom
\section{Introduction}

It is well-known that the framework of a TeV-scale composite Higgs (CH) addresses the Planck-weak hierarchy problem of the standard model (SM) (see e.g.~\cite{Panico:2015jxa} for a review). At the same time, this idea can be probed, either directly by producing accompanying TeV-scale composites at the LHC and future colliders or 
indirectly via virtual effects of such compositeness on properties of the SM particles (including the Higgs boson). Indeed, both kinds of searches are already leading to significant constraints on the Higgs compositeness scale.

Remarkably, Higgs compositeness coupled with partial compositeness of the SM fermions also accounts for the flavor puzzle of the SM, i.e., hierarchies in the masses and mixings of the SM fermions as follows \cite{Kaplan:1991dc,Gherghetta:2000qt,Agashe:2004cp, Contino:2004vy}. %
The
central idea is to have elementary\footnote{``Elementary'' here means ``external to the composite sector''. In this paper we will use both, elementary and external, interchangeably.} massless fermions mix with massive Dirac composites, with the SM fermions being the resulting massless admixtures.
The key point is that this mixing is controlled by the anomalous dimensions of the operators of the strong sector involved in this mixing, which can be realized if the strong dynamics is approximately conformal over a large hierarchy of scales, such as the Planck-weak hierarchy.
Hence, modest variations in the scaling dimensions of these conformal field theory (CFT) operators can result in large hierarchies in the degree of compositeness of SM fermions. 
Now, the coupling of the SM fermions to the Higgs proceeds via the composite component of the SM fermions because the Higgs is fully composite. This then results in hierarchical SM Yukawa couplings, thus also the SM fermion masses. 

While the above is more-or-less the full story for {\em charged} SM fermions, the case of the SM neutrinos is  more nuanced as follows. To begin with, Dirac masses for neutrinos proceed in the same way as for the charged fermions, i.e., arising from partial compositeness of {\em both} the left-handed (LH) neutrino --- which is part of the lepton $SU(2)_L$ doublet --- and the SM gauge singlet right-handed (RH) neutrino $N$. 
However, we then need to consider the fate of lepton-number: we assume here that CFT respects it, but the  elementary sector and its mixing with the strong sector break  it, thus a high-scale Majorana mass for the elementary $N$, denoted by 
$m_N \gg$ TeV is allowed.
Since the heavy Majorana $N$ mixes with composite singlet neutrinos $\Psi$, integrating out $N$ generates a tiny Majorana mass terms for $\Psi$, which constitute an entire {\em tower} of pseudo-Dirac states starting at $\sim$ TeV.
The final SM neutrino mass is generated by exchange of these pseudo-Dirac composite SM singlet neutrinos. To sum up, what started-out looking like high-scale usual type I seesaw mechanism morphs into inverse seesaw~\cite{Agashe:2015izu}!

Such a seesaw
scenario
for Majorana SM neutrino mass naturally motivates studying baryon asymmetry of the universe (BAU) via leptogenesis~\cite{Fukugita:1986hr}.
This idea is especially even more interesting in the present model due to sort of {\em combination} of {\em two}-scale processes: those involving decays of heavy Majorana $N$, akin to usual high-scale seesaw leptogenesis, and dynamics of TeV scale
pseudo-Dirac $\Psi$, which is the inverse seesaw component.

Of course, leptogenesis has been extensively analyzed for high-scale type I seesaw as well as inverse seesaw. 
However, to the best of our knowledge, simultaneous presence of both these phenomena in the same framework is a unique feature built into the CH model, and no comprehensive study of generation of BAU was carried out. The goal of this paper is two-fold: (i) to develop a formalism and (ii) to study physics of baryogenesis in CH models in detail. We dub this scenario of leptogenesis in nearly conformal sectors (like CH models) as \emph{Conformal Leptogenesis}.

In fact, in previous works~\cite{ Agashe:2018oyk, Agashe:2018cuf}, we initiated the study of this conformal leptogenesis by proposing a simplified model for the analysis of such a scenario by restricting to just $N$ and the {\em lightest} $\Psi$ {\em at} $\sim$ TeV and treating the latter as regular particles even at the high-temperatures relevant for leptogenesis.
We showed that the lepton asymmetry from the inverse seesaw module, i.e., decays of $\Psi$ into SM particles, is generically not enough, so we were led to 
consider also decays of heavy $N$ into $\Psi$: see refs \cite{ Agashe:2018oyk, Agashe:2018cuf} for details.
This ``UV'' asymmetry can be large, but it is still ``modulated'' by TeV-scale physics, for example, washout processes due to $\Psi$. 
The observed BAU can be acheived in this model with the interesting interplay between high and low-scale physics.

In this paper, we demonstrate that the above toy model can in fact be promoted to the fully realistic conformal leptogenesis, albeit we need overcoming of several challenges en-route to such a fully-working version.
The crucial point is that $\Psi$ are {\em composites} arising from the confinement of the strongly coupled sector at around TeV scale. Thus, at temperatures $T\gg$ TeV, which are necessarily involved in dealing with the lepton asymmetry generation from decays of heavy elementary $N$, the composite sector is in the deconfined phase and thus these $\Psi$ do {\em not} exist. In this case, $N$ decays into  $\Psi$ constituents or CFT ``{\em un}particle'' states \cite{Georgi:2007ek}. 
Therefore, we first have to develop the formalism for the generation of asymmetry from $N$ decays in such a scenario, including washout effects.
One immediate challenge is that the usual Boltzmann equations formulated in terms of particle number distribution functions are not applicable, as the CFT does not admit a particle description. One then needs to write the equations in terms of well defined CFT correlation functions. 
As we will discuss, this is possible by considering the Boltzmann equations as the effective description of the dynamics of the slowly varying expectation values of the ``charge'' of the CFT as well as the charges of the elementary sector. 
In fact, this approach has been already applied to the standard type-I seesaw leptogenesis. In ref.~\cite{Bodeker:2014hqa} such a formalism was developed as an effective description of small and slowly varying deviations from equilibrium. They expressed the washout coefficients in terms of the SM correlation function in equilibrium. Later in ref.~\cite{Bodeker:2017deo} this was extended to include the source term in the nonrelativistic regime. This formalism allowed them to systematically compute  the higher order corrections to these coefficients from SM couplings.
Applying the same approach, we obtain Boltzmann equations for conformal leptogenesis that have a similar form, as a consequence of the ``universality'' of the effective description, while the coefficients can be computed in terms of the CFT correlation functions.

The next stages are also strikingly non-trivial in conformal leptogenesis, starting with the
subsequent
evolution of this lepton asymmetry created at temperatures $\gg$ TeV down to $\sim$ TeV 
where confinement occurs: once again, the standard treatment, for e.g. done for simplified model, is not quite sufficient due to the CFT sector being deconfined. 
Indeed, we think that our above
theoretical 
analysis of asymmetry generation and evolution in a CFT can be applied more generally, not just to leptogenesis. 

Then, we have to consider the impact of the actual phase transition (PT) leading to confinement at $\sim$ TeV, including possible dilution of asymmetry due to a ``supercooling'' effect: this phenomenon has already been studied in a general context (see reference~\cite{Creminelli:2001th, Randall:2006py, Nardini:2007me, Konstandin:2010cd, Konstandin:2011dr, Bunk:2017fic, Baratella:2018pxi,Bruggisser:2018mrt, Megias:2018sxv, Agashe:2019lhy,Fujikura:2019oyi, DelleRose:2019pgi,Agashe:2020lfz, Ares:2020lbt,Agrawal:2021alq, Levi:2022bzt, Csaki:2023pwy,Eroncel:2023uqf,Mishra:2023kiu,Mishra:2024ehr}).
Obviously, this segment of the picture did not appear 
at all in the simplified case. 
Finally, there can be additional washout and related effects after the PT originating from TeV-scale composites $\Psi$: this part is more-or-less similar to that in the simplified model, with one main difference. The difference arises from the fact that after the PT where composites form, the temperature tends to be below the mass of the $\Psi$, therefore leading to suppression of the inverse decay rates by a Boltzmann factor $\sim e^{-m_{\Psi}/T}$ (similar effect has also been pointed out in ref.~\cite{Dichtl:2023xqd}, where they studied leptogenesis from decay of resonances from a composite sector).

Combining all these interesting aspects of conformal leptogenesis, we show that observed BAU can be indeed achieved in a broad range of parameter space.
It is noteworthy that successful conformal leptogenesis requires a sizeable degree of compositeness of SM neutrino. We show that this feature is consistent with charged lepton mass hierarchies and current precision tests in the electroweak (EW) and flavor/CP violation sectors. Moreover, it has the potential to be detected in the future experiments like FCC-ee with a more precise measurement of invisible decay width of SM $Z$ boson.

The outline of the rest of this paper is as follows. In section~\ref{sec:neutrinomass}, we review seesaw mechanism and generation of neutrino masses in CH models. In section~\ref{sec:HighscaleBE}, we derive Boltzmann equations governing the dynamics of the asymmetry before the confinement. Based on these Boltzmann equations, we discuss in section~\ref{sec:evolution_asymmetry_high_scale} the generation and evolution of the asymmetry at the high scale. The low scale dynamics involving the effects of the confinement phase transition and washout from composite resonances are discussed in section~\ref{sec:low_scale_dynamics}. Combining all the above effects, the observed BAU is obtained and other relevant phenomenology is discussed in section~\ref{sec:baryon_asymmetry}. We then conclude in section~\ref{sec:conclusion}. In appendix~\ref{app:nearequilibrium}, we detail the derivation of the structure of Boltzmann equations used in the main text, and discuss the possible finite temperature corrections in appendix~\ref{app:finiteT}. We also present an intuitive derivation of the Boltzmann equations in appendix~\ref{app:moreintuitivederivation}.

\section{Review of Neutrino Masses in Composite Higgs Models} \label{sec:neutrinomass}

In this section we review the generation of neutrino masses in CH models implementing partial compositeness~\cite{Kaplan:1991dc,Gherghetta:2000qt,Agashe:2004cp, Contino:2004vy}. 
We first briefly recall how the flavour structures for quarks and charged leptons are obtained, and then discuss a few scenarios for neutrino masses, specializing on the model we consider in this work.

The basic assumptions underlying partial compositeness are the following. The theory contains a strongly coupled sector that is approximately conformal from the UV cutoff $\Lambda$ down to a scale $m_\rho$ where resonances, including a CH, appear. For simplicity we assume that $m_\rho$ and $g_\rho$ denote the typical mass and coupling of these resonances, and that the strong dynamics has correlation functions that can be estimated using naive dimensional analysis up to numbers of order unity. We will refer to this dynamics as the ``CFT'' or the ``strong sector'', since by assumption $g_\rho$ will always be the larger coupling involved. In addition to the strong sector, there exists a ``weak'' or ``fundamental'' sector composed of the SM fermions and gauge fields. The two sectors couple via weak gauging of the SM gauge group and via a mixing between the SM fermions and composite fermionic operators of the strong sector. The latter is the only source of breaking of the accidental flavor $U(3)$'s of the fundamental fermions. Specifically, each elementary fermion field, $f_i$, couples to a fermionic operator $O_{f\alpha}$ of the strongly coupled CFT sector via a linear coupling 
\beq
\mathcal{L} \supset -\sum_{f} [\lambda_{f} ]_{i\alpha} f_i O_{f \alpha} + \textrm{h.c.}, \label{eq:fermion_couplings}
\eeq
where only left-handed Weyl fermions are used, $i,j=1,2,3$ are the generation indices of elementary fermion fields and $\alpha,\beta=1,2,3$ are those of the composite sector, and $\lambda_{f}$ denotes the coupling. Because in this paper we will mainly deal with leptons it is useful to write eq.~\eqref{eq:fermion_couplings} explicitly 
\beq\label{PCchargedleptons}
\mathcal{L} \supset -[\lambda_{L}]_{i\alpha} L_i O_{L \alpha} -[\lambda_{e}]_{i\alpha} e^c_i O_{e \alpha}+ \textrm{h.c.},
\eeq 
with $L_i$ and $e^c_i$ denoting the $SU(2)_L$ lepton doublets and  singlets respectively.

Let us now see how the charged leptons acquire a Yukawa coupling in this framework. Similar considerations apply to quarks. At the confinement scale $m_\rho$ the coupling $\lambda_f f O_f$ is mapped into a mass mixing between the SM fermions and resonances of the strong sector. The light state that results from the mixing has a compositeness of order $\sim \bar{\lambda}_{f}/{g_\rho}$, where $\bar\lambda_f(\mu)$ denotes the renormalized and dimensionless version of $\lambda_f$, defined explicitly in eq.~\eqref{eq:lambda_bar} below. By construction the resonances couple to the Higgs with a coupling $\sim g_\rho$. So, via the mixing the SM fermions inherit a coupling of order $(\bar{\lambda}_{f}/{g_\rho})(\bar{\lambda}_{f}/{g_\rho})g_\rho$. More precisely, this picture leads to the following Yukawa 
\beq
y_{ij}= [\bar{\lambda}_{L}(m_\rho)]_{i\alpha} [\bar{\lambda}_{e} (m_\rho)]_{j\beta}\frac{c_{\alpha\beta}}{ g_\rho},
\label{eq:Yukawa_couplings}
\eeq
where $c_{\alpha\beta}$ are unknown numerical coefficients of the strong dynamics. Even starting with flavor anarchic couplings $\bar\lambda_f(\Lambda)$ in the UV, a desired hierarchical flavor structure in the IR may be achieved via renormalization group flow since a mild difference in the scaling dimensions of $O_{f \alpha}$ can lead to large variations in the IR couplings $\bar\lambda_f(m_\rho)$.

We next turn to the neutrinos. There are several ways to generate neutrino masses in such a setup \cite{Frigerio:2018uwx}. For example, one option is to assume that the strong sector contains a $SU(2)_L$-triplet scalar operator  $\mathcal{O}_{ T}$ carrying no lepton number. In this case one can write a coupling $c_{ij} L_i L_j \mathcal{O}_T$ that, provided ${\cal O}_T$ has a non-trivial matrix element with two CHs, generates a Majorana mass matrix for neutrinos with generic texture if $c_{\alpha \beta}$ is anarchic. Small neutrino masses are obtained if the dimension of $\mathcal{O}_T$ is sufficiently large \cite{Keren-Zur:2012buf}. 

We will, however, consider another possibility. We introduce a set of elementary left-handed SM-singlet Weyl fermions $N_i$, with non-vanishing Majorana masses, which couple to $SU(2)_L$-singlet fermion operators $ O_\alpha$ of scaling dimension $\Delta_\alpha$
 via linear couplings, similar to the charged leptons in eq.~\eqref{PCchargedleptons}: 
\beq
{\cal L}  \supset  N^\dagger_{i} i \overline{\sigma}^{\mu}\partial_{\mu}N_{i}-  \frac{1}{2}m_{N_{i}} N_{i}N_{i}-\lambda_{i\alpha} N_{i} O_{\alpha}+\textrm{h.c.} 
\label{eq:B-L_breaking}
\eeq
We further assume that the strong sector possesses an accidental global $U(1)$ which we identify as lepton number, under which $O$ has unit charge. 
Now, in the limit $m_{N_i} \rightarrow 0$ a conserved lepton number may be defined in the full composite-fundamental system, and in principle we may realize Dirac neutrino masses in analogy with the charged leptons and quarks. In this work, instead, we will focus on the case where the Majorana masses $m_{N_i}$ are large compared to $m_\rho$, a regime which generates tiny Majorana neutrino masses through the seesaw mechanism as we will discuss next. Such a possibility was considered as a realization of type-I seesaw in RS setup~\cite{Huber:2003sf}, and later its CFT dual description was given in \cite{Agashe:2015izu}. In particular, despite its resemblance to the type-I seesaw Lagrangian, it was shown in \cite{Agashe:2015izu} through both 5D and its dual 4D analysis that this model is actually a realization of a form of inverse seesaw.\footnote{For resulting LHC phenomenology, see \cite{Agashe:2016ttz,Agashe:2017ann}.}

In the remainder of this section we will review how neutrino masses are generated in the setup of eq. \eqref{eq:B-L_breaking}. We will emphasize the existence of various scenarios depending on the scaling dimensions $\Delta_\alpha$ of the CFT operators $O_\alpha$ and of the size of the couplings involved. In any unitarity CFT $\Delta_\alpha$ must satisfy $\Delta_\alpha \geq 3/2$. Yet for $3/2<\Delta_\alpha<2$, and in the large-$N_c$ regime considered here, the strong sector possesses a relevant deformation $O^2$ that breaks the lepton number. This violates our hypothesis that the strong sector is an approximate CFT with accidental lepton number. We will thus discard this possibility and stick to the regime $\Delta_\alpha>2$. Our discussion will be qualitative; for this reason we will momentarily suppress the flavor indices and analyze a simplified one-flavor scenario. We will restore the flavor indices in the subsequent sections.

\subsection{$\Delta > 5/2$}

Let us first consider scenarios with $\Delta>5/2$. In this case the interaction term $N O$ is irrelevant, which means that it is suppressed in the IR. To see this, observe that we can conventionally rescale the operator $O$ so that its engineering dimension coincides with the scaling dimension, $\Delta$. Then the coupling $\lambda$ in \eqref{eq:B-L_breaking} must have a mass dimension equal to $5/2-\Delta$. What measures its strength in a generic physical process computed at some energy scale $\sim\mu$ is the dimensionless combination \begin{eqnarray}\label{eq:lambda_bar}
    \bar\lambda(\mu)\equiv \lambda \,\mu^{\Delta -5/2}.
\end{eqnarray}
We then see that with $\Delta>5/2$, the smaller the energy scale $\mu$ the smaller the dimensionless coupling: we say that such coupling is irrelevant at low energies. Intuitively, we expect the neutrino mass to be controlled by the IR coupling
\bea
\bar\lambda(m_\rho)=\bar\lambda(\Lambda) \left(\frac{m_\rho}{\Lambda}\right)^{\Delta -5/2}
\eea
because the relevant scale at which the neutrino mass operator is generated is $m_\rho$. We then anticipate that the neutrino mass can be tiny even if $\bar\lambda(\Lambda)$ is taken to be sizable as long as $m_\rho\ll\Lambda$. The estimate provided below will confirm this expectation.

We approach the problem using a Wilsonian picture, starting from the Lagrangian in eq.~\eqref{eq:B-L_breaking} at the UV cutoff and asking what EFT is generated at the confinement scale $m_\rho$. To start, we observe that under our assumption $m_N\gg m_\rho$ the strong sector remains an approximate CFT when we decrease the Wilsonian scale down to the $m_N$ threshold. Furthermore, because of our convention that the engineering and scaling dimensions coincide, the operator $O$ has no scale-dependence, i.e. no anomalous dimension. Hence there is no non-trivial RG effect in going from $\Lambda$ to $m_N$: the EFT at $\mu>m_N$ basically coincides with eq.~\eqref{eq:B-L_breaking}. At the threshold $\mu\sim m_N$ we integrate out $N$ and obtain an EFT defined by an approximate CFT perturbed by an irrelevant operator $O^2$.

Decreasing $\mu$ further we encounter a new mass threshold at $m_\rho\ll m_N$. We will assume the dimension of $O^2$ is $2 \Delta$.\footnote{However, in general this may not be the case. For the latest bounds from conformal bootstrap applicable to this case see e.g. figure 7 of \cite{Karateev:2019pvw}. For example an operator $O$ of dimension $\Delta=1.9$ (so a relevant coupling $\lambda$) could still yield $O^2$ of dimension as large as $7$. On the other hand, it is also possible that the deformation $\lambda$ is irrelevant with for example $\Delta=2.7$ and runs to become very small at scale $m_N$, but then the operator $O^2$ could have dimension considerably smaller than $2 \Delta$, compensating for the smallness of $\lambda$.}  This holds if the CFT sector is described by a large-$N_c$ gauge theory. By an argument completely analogous to the one given above, our deformation $O^2$ has no anomalous dimension; its running from $m_N$ all the way to $m_\rho$ is trivial. The effective field theory at $m_\rho$ is thus schematically given by
\beq\label{eq:CFTatmrho}
{\cal L}({m_\rho})\supset {\cal L}_{\rm CFT}+\left[\frac{\lambda^2}{2m_N}O^2+{\text{h.c.}}\right].
\eeq

At the scale $m_\rho$
 the operator $O^2$ should finally match onto EFT operators carrying two units of lepton number and involving only SM fields. The natural expectation is that the dominant such operator be the dimension-5 Weinberg operator
 \bea\label{matchingOHL}
 O^2\to s_\nu^2m_\rho^{2\Delta-5}({\cal H}L)^2.
 \eea
The factor of $m_\rho^{2\Delta-5}$ is simply based on dimensional analysis and $s_\nu$ is a suppression factor that measures the overlap of $O$ with the ``neutrino component'' ${\cal H} L$ where ${\cal H}$ refers to the SM Higgs doublet. In minimal models $s_\nu$ coincides with the parameter $\bar\lambda_L(m_\rho)/g_\rho$ controlling the mass matrix of the charged leptons. Yet, those two quantities are not the same in more general scenarios where the fundamental lepton doublets mix with different operators of the strongly coupled sector (see for example \cite{Agashe:2009tu}).\footnote{ \label{footnote:snusL} In more detail, the strong sector/CFT respects an extended EW {\em global} symmetry $SU(2)_L \times SU(2)_R \times U(1)_X$, of which only the SM subgroup, i.e., $SU(2)_L \times U(1)_Y$ is gauged, where
$Y = T_{ 3 R } + X$: this full selection rule applies also to couplings of 
Higgs to the CFT operators.
However, the couplings of elementary fermions to the CFT need to be invariant only under the (gauged) SM symmetry.
So, we can choose, for example, ${O}_{L}$ 
to transform -- under $SU(2)_L \times SU(2)_R \times U(1)_X$ -- as $(2,2,0)$, while ${O}_{\nu} \sim (2, 1, -1/2)$: the lower $SU(2)_R$-component of ${ O}_L$ and ${ O}_{ \nu }$ 
are {\em both} then $(2, -1/2)$ under $SU(2)_L \times U(1)_Y$. Hence, the elementary $L$ can couple to both ${ O}_L$ and ${ O}_{ \nu }$. As far as $SU(2)_L$ singlet sector is concerned, we choose 
elementary $e$ to couple to lowest $SU(2)_R$ component of ${ O}_{e} \sim (1, 3, 0)$ in such a way that charged lepton coupling to Higgs -- which is $(2,2, 0)$ -- has to proceed via
$\lambda_{L}$ coupling of elementary $L$: the point is that ${O}_{e} \cdot  { O}_L \cdot \hbox{Higgs}$ then forms an $SU(2)_L \times SU(2)_R \times U(1)_X$ singlet as required. 
Similarly, we can choose elementary $N$ to couple to [$SU(2)_R$ upper-component of] ${O} \sim (1, 2, -1/2)$ so that neutrino coupling to Higgs goes through $\lambda_{ \nu  }$ coupling of elementary $L$, with 
$\hbox{Higgs} \cdot {O }_{ \nu } \cdot { O}$ again being a singlet.
Alternatively, we can choose ${ O}_L \sim (2, 1 , -1/2)$ and ${ O}_{\nu } \sim (2, 2, 0)$ [using the lower 
$SU(2)_R$-component thereof], 
while 
${ O}_{e} \sim (1 , 2, -1/2)$ and ${ O} \sim (1,1,0)$.
} In this work we will in fact consider the more general possibility $s_\nu \neq \bar\lambda_L/g_\rho$ since, as we will see, successful leptogenesis imposes important constraints on $s_{\nu}$, which for $s_\nu=\bar\lambda_L/g_\rho$ would not always be compatible with existing bounds from observables involving the charged leptons.

The neutrino mass resulting from our model is finally obtained combining eqs.~\eqref{eq:CFTatmrho} and \eqref{matchingOHL}. After the Higgs has acquired its vacuum expectation value 
$v=174$ GeV, we have
\beq
m_\nu \sim s_{\nu}^2 \frac{\lambda^2}{m_N}m_\rho^{2\Delta-5}v^2\sim s_{\nu}^2\left(\frac{m_\rho}{\Lambda}\right)^{2\Delta-5} \frac{\bar\lambda^2(\Lambda)v^2}{m_N}\sim s_{\nu}^2\left(\frac{m_\rho}{m_N}\right)^{2\Delta-5} \frac{\bar\lambda^2(m_N)v^2}{m_N}.
\label{eq:m nu gtr 25}
\eeq
The final expression takes the same form found in the familiar seesaw scenario, though here it is modulated by an RG running factor $\left({m_\rho}/{\Lambda}\right)^{2\Delta-5}=\bar\lambda^2(m_\rho)/\bar\lambda^2(\Lambda)$. The latter can be significantly smaller than one if $NO$ is strongly irrelevant. Compared to the standard type-I seesaw, our model opens more parameter space of $\bar\lambda(\Lambda)$ and $m_N$ that can obtain the observed neutrino masses.

\subsection{$2 < \Delta < 5/2$}

We next consider scenarios with $2<\Delta<5/2$. Now the operator $N O$ is relevant and so the coupling $\bar\lambda$ grows in the IR. Nevertheless, $O^2$ remains irrelevant. In the large-$N_c$ limit, the running of $\bar\lambda$ follows the RG equation (see e.g.~\cite{Contino:2004vy}) 
\beq \label{eq:RG for lambda}
\mu \frac{d \bar\lambda}{d \mu} = \left( \Delta - \frac{5}{2} \right) \bar\lambda + b \frac{N_c}{16\pi^2} \bar\lambda^3 + \cdots ,
\eeq
where $b>0$ is a numerical coefficient of order unity and the dots refer to higher order corrections in $\bar\lambda$. Ignoring the latter, one finds an IR fixed point $\bar\lambda_* = \sqrt{{(5/2-\Delta)}/{b}}\;{4\pi}/{\sqrt{N_c}}$. Actually, the coupling is well approximated by the fixed point value for all scales $\mu\leq\Lambda_* \sim \Lambda  (\bar\lambda(\Lambda)/\bar\lambda_*)^{\frac{1}{5/2-\Delta}}$. We will assume that the existence of the IR fixed point persists once the higher order corrections in $\bar\lambda$ are taken into account.\footnote{The existence of the fixed point and its value can be trusted only if $\Delta-5/2$ is small. If not,  the higher order corrections in $\bar{\lambda}^2 N_c/(16 \pi^2)$ can in general be important and change the value of the fixed point or spoil its existence altogether. In this subsection we assume that the fixed point still persists, especially given that $|\Delta-5/2|<0.5$ may lead to a small enough $\bar{\lambda}_*$. But we will however avoid this regime for our leptogenesis analysis. } In addition, for simplicity we will take the transition between quickly running coupling (at $\mu>\Lambda_*$) and slowly running coupling (at $\mu<\Lambda_*$) to be basically instantaneous. One can then envision two possibilities.

The first possibility corresponds to having $m_N>\Lambda_*$. Starting with a small $\bar\lambda(\Lambda)$, the coupling remains perturbatively small down to the scale $\mu=m_N$. The field $N$ is always weakly-coupled and we can integrate the associated particle out as done in the case $\Delta>\frac{5}{2}$. In this situation one obtains the same effective field theory and parametrically the same expression for the neutrino mass in eq. \eqref{eq:m nu gtr 25}. There is of course a quantitative difference: the RG factor $\left({m_\rho}/{\Lambda}\right)^{2\Delta-5}=\bar\lambda^2(m_\rho)/\bar\lambda^2(\Lambda)$ corresponds now to an enhancement, so a realistic $m_\nu$ requires either a larger $m_N$ or a very small $\bar\lambda(\Lambda)$.

The second possibility is realized when $\Lambda_*>m_N$. In this case either $\bar \lambda\sim\bar\lambda_*$ in the UV or $\Delta$ and $m_N$ are such that $\bar \lambda$ grows to a point where the two terms appearing in eq.~\eqref{eq:RG for lambda} become comparable at $\mu > m_N$. Now the situation is completely different from the cases mentioned earlier. Qualitatively, the coupling $\bar \lambda$ increases from $\Lambda$ to $\Lambda_*$ and stops running below $\Lambda_*$, at least approximately. In other words, the operator $N O$ is relevant above $\Lambda_*$ and marginal below that scale. Yet, because $\bar{\lambda}_*$ is small in the large-$N_c$ limit, the effect of the Majorana fermion $N$ on the CFT is negligible, e.g. the dimension of $O$ remains close to the value $\Delta$ of the CFT in isolation. On the other hand, the CFT modifies the dynamics of $N$ at order $\bar\lambda^2 N_c/16\pi^2$, which is a sizable effect at the fixed point. This has important consequences. The most important one is that the scaling dimension of $N$ can be, and indeed does get, corrected at $\mathcal{O}(1)$. More explicitly, since $N O$ becomes marginal and the scaling dimension of $O$ remains the same, we conclude that $N$ must acquire a scaling dimension $4-\Delta$. The field has an anomalous dimension $\gamma_N=4-\Delta-3/2=5/2-\Delta$ that can be sizable. Our original weakly-coupled Weyl field has thus turned into a strongly-coupled operator.

To avoid confusion we stress that because the engineering and scaling dimensions of $N$ do not coincide, the marginal operator $NO$ appears in the EFT with a scale-dependent coefficient. That is, the running from $\Lambda_*$ to $\mu<\Lambda_*$ must take into account the effect of the non-trivial $\gamma_N$, so that the interaction $\lambda NO=\bar\lambda(\Lambda_*)\,\Lambda_*^{5/2-\Delta}NO$ at lower scales becomes: 
\bea\label{renofON}
{\cal L}(\mu\lesssim\Lambda_*)\supset -\bar\lambda(\Lambda_*)\,\Lambda_*^{5/2-\Delta}\left(\frac{\mu}{\Lambda_*}\right)^{\gamma_N}NO=-\bar\lambda(\Lambda_*)\,\mu^{5/2-\Delta}NO.
\eea
This is nothing but a rephrasing of the previous statement that the dimensionless coupling $\lambda\mu^{\Delta-5/2}$ does not run.

Another significant effect of the non-trivial fixed point is seen in the mass operator $N^2$. In our large-$N_c$ framework $N$ has a 2-point function but suppressed higher point correlators. As a result, the scaling dimension of the mass operator $N^2$ is approximately twice the one of $N$, namely $8-2\Delta$. More explicitly, the anomalous dimension of the mass operator $N^2$ is approximately $2\gamma_N=(8-2\Delta)-3=5-2\Delta$ and satisfies $0<2\gamma_N<1$ in the regime of interest. The operator has a dimension that is larger than in the free theory but nevertheless is always relevant. Then, slightly below $\Lambda_*$ the mass interaction in eq.~\eqref{eq:B-L_breaking} runs according to
\bea\label{renofNN}
&& \mathcal{L}({\mu\lesssim\Lambda_*}) \supset -\frac{m_N}{2} \left( \frac{\mu}{\Lambda_*} \right)^{5-2\Delta} NN.
\eea
Eventually, at a scale we denote $m_{N*}$, the mass operator gets so strong that the approximate conformality of the dynamics involving $N$ maximally breaks down. But as long as we stick to the large $N_c$ limit, such a breaking has only a small impact on the original ``strong'' CFT. Instead, what we expect to happen at the strong-coupling regime $\mu\sim m_{N*}$ is that heavy resonances of mass $m_{N*}$ associated to the strong $N$ dynamics start to decouple. The dynamics below that scale is approximately described by the usual $O^2$ operator as in our previous examples. The main difference is that its coefficient is obtained by integrating out the singlet states at a different scale $m_{N*}\neq m_N$. 

With this in mind, our EFT at $m_\rho$ is thus found analogously to the case $\Delta>5/2$. Using eqs.~\eqref{renofON} and \eqref{renofNN}, the Lagrangian in eq.~\eqref{eq:B-L_breaking} renormalized down to $m_{N_*}$ gives
\bea\label{reno}
&& \mathcal{L}(m_{N*}) \supset -\bar\lambda(\Lambda_*)\,m_{N*}^{5/2-\Delta}NO-\frac{m_{N*}}{2} NN+{\text{h.c.}}.
\eea
Integrating out the massive modes of $N$ the EFT at $m_\rho$ finally is
\begin{align}\label{EFTDlfh}
{\cal L} (m_\rho)
&={\cal L}_{\rm CFT}+\left[\frac12\bar\lambda^2(\Lambda_*)m_{N*}^{4-2\Delta} O^2+{\text{h.c.}}\right]\\\no
&={\cal L}_{\rm CFT}+\left[\frac{\lambda^2}{2m_{N*}}\left(\frac{m_{N*}}{\Lambda_*}\right)^{5-2\Delta} O^2+{\text{h.c.}}\right],
\end{align}
where we used $\bar\lambda(\Lambda_*)=\lambda\,\Lambda_*^{\Delta-5/2}
$ to write the second line in terms of $\lambda$. To compare to the previous result shown in eq.~\eqref{eq:CFTatmrho}, found for $\Delta>5/2$ or for $2<\Delta<5/2$ with $\Lambda_*<m_N$, we have to estimate $m_{N*}$.

We can find a benchmark for the mass scale $m_{N*}$ via a simple scaling argument. To better appreciate it, it is convenient to first rephrase in our language what is the relevant scale in the familiar case of a free fermion of mass $m_N$. The scale where a weakly-coupled fermion is integrated out is the one where the associated dimensionless coupling $m_N/\mu$ becomes of order unity and thus cannot be treated anymore as a perturbation. Similarly, the scale at which we are to integrate out the strongly-coupled states of $N$ can be identified with the mass scale $m_{N*}$ at which the dimensionless coupling $\bar m_N$ associated to the mass perturbation $N^2$ can no longer be considered small. Inspecting eq.~\eqref{renofNN} we see that for $\mu<\Lambda_*$ the dimensionless coupling associated to $N^2$ should be defined as
\beq\label{runmassN}
\bar m_N(\mu)\equiv\frac{m_N}{\mu}\left(\frac{\mu}{\Lambda_*}\right)^{5-2\Delta}.
\eeq
The mass scale $m_{N*}$ we are seeking is implicitly defined by the condition $\bar m_N(m_{N*})\sim1$, or equivalently
\beq\label{mNstar}
\frac{1}{m_{N*}}\left(\frac{m_{N*}}{\Lambda_*}\right)^{5-2\Delta}\sim\frac{1}{m_N}.
\eeq
From this expression it is easy to see that $m_{N*}> m_N$, consistently with it being a relevant deformation. But the most striking implication is actually that plugging eq.~\eqref{mNstar} into eq.~\eqref{EFTDlfh} the effective field theory becomes parametrically the same as eq.~\eqref{eq:CFTatmrho}, despite the physics leading to it is a priori qualitatively different. As a consequence, also the neutrino mass is given by the same parametric expression. An explicit confirmation of the parametric equivalence between the neutrino masses in the two regimes $\Delta>5/2$ and $2<\Delta<5/2$ can be found in the five-dimensional scenario studied in \cite{Agashe:2015izu}, where it is seen that the five-dimensional result is unaffected when $\Delta$ passes across the threshold $5/2$.

We have completed our review. In the following we will focus on either the scenario $\Delta>5/2$ or the scenarios $2<\Delta<5/2$ with $\Lambda_*\ll m_N$ because our leptogenesis analysis relies on the assumption that $N$ be weakly-coupled. Whether or not our conclusions can be extended to the regime $\Lambda_*>m_N$, where we have seen $N$ acquires a large anomalous dimension, is an interesting question that needs further investigation.

\section{Boltzmann Equations above the Confinement Scale}
\label{sec:HighscaleBE}

Having reviewed the theory of neutrino masses and mixings in CH models, we now move to a discussion of leptogenesis in the setup. 
We will assume that the CFT and the interactions in eq.~\eqref{eq:fermion_couplings} perturbatively conserve the SM baryon $B$ and lepton numbers $L$, while  non-perturbatively,  sphalerons violate both of them but preserve $B-L$.  $B-L$ is thus only broken by the interactions in eq.~\eqref{eq:B-L_breaking} with simultaneously nonzero $m_{N}\neq0$ and $\lambda\neq 0$. If during the generation of the asymmetry the sphalerons are out of equilibrium, then the decays of $N$ directly generate a lepton asymmetry in $-L$, which is eventually converted into a $B-L$ charge once the $(B+L)$-violating sphaleron interactions get into equilibrium. Alternatively, it is possible that during the asymmetry generation the sphaleron interactions are already in equilibrium. Under such circumstance, the asymmetry generated from the decays of $N$ is already in the form of a $B-L$ charge.\footnote{Ref.~\cite{Garbrecht:2014kda} estimated that the EW sphaleron processes in a universe dominated by the SM radiation bath gets into equilibrium below $T \sim 10^{12}$ GeV.}
Irrespective of which case is actually realized, in this section we will keep referring to the production of a ``lepton asymmetry'' and denote the associated quantum number as $n_L$ or $Y_L$.

In this section we introduce the basic ingredients necessary to compute the lepton asymmetry down to temperatures larger than the confinement scale $m_\rho$. Explicit estimates for the present-day asymmetry must also take into account the phase transition at $m_\rho$ and will be discussed in subsequent sections.
We begin in section \ref{subsec:BELV} by presenting the basic Boltzmann equations that govern the generation and depletion of the asymmetry at temperatures $T\gg m_\rho$, as well as by discussing the main underlying assumptions. When these equations apply, the model-dependence is captured by a few parameters that can be expressed in terms of correlators of the strong dynamics, and that can be calculated explicitly in the large $N_c$ approximation as shown in section \ref{subsec:computeepsilonGamma}. Details on the derivation of the relevant Boltzmann equations, as well as more intuitive arguments justifying them, are provided in the appendices. The most important result of this section is that there exists a significant portion of the parameter space of our models in which simple equations provide an accurate description of the physics, up to small corrections suppressed by powers of $T/m_N$ in the non-relativistic regime.

\subsection{Boltzmann Equations } 
\label{subsec:BELV}

In this subsection we present the equations governing the production and evolution of the lepton asymmetry at $T\gg m_\rho$. 
At first sight, one may think that ordinary Boltzmann equations constructed in terms of distribution functions of \emph{particles} cannot be justified in our models. This is because our framework involves a strongly coupled, approximately conformal field theory sector at finite temperature, which in general does not admit a description in terms of particle s. 
However, as we argue next, the set of equations that are relevant to the production and evolution of the lepton asymmetry can be written without any reference to hypothetical particle distribution functions for the strong sector. 

The basic observation is that the key players in leptogenesis are a set of ``slowly varying densities'' that are not directly linked to the nature of the CFT constituents. They are the number density $n_N$ of the singlets $N_i$ and the lepton number density $n_L$. As long as these quantities vary slowly with respect to their equilibrium values ($n_N^{\text{eq}}$ and $0$ respectively) on long distances and times (compared to the time and length scales of the microscopic theory as well as those set by the temperature), an effective description for them can be provided in a rather \emph{universal} fashion. Not surprisingly, the resulting set of equations is similar to those for the familiar Type-I seesaw leptogenesis. These basic equations, to be justified shortly, are: 
\beq \label{eq:BoltzmanneqNLV}
\frac{d n_N}{d t} + 3 H n_N = -\langle\Gamma_N\rangle_T \left( n_N- n^{\rm eq}_N (T) \right),
\eeq
\beq \label{eq:BoltzeqAsymLV}
\frac{d n_L}{d t} + 3 H n_L \approx -\frac{6 }{ c}  \, \frac{n_N^{\rm eq}(T)}{T^3} \langle\Gamma_N\rangle_T \, n_L+ \epsilon \,  \langle\Gamma_N\rangle_T \left( n_N- n^{\rm eq}_N (T) \right),
\eeq
where the Hubble parameter is
\beq
H = \frac{d\ln a}{dt} =\sqrt{\frac{8\pi \rho}{3m_{\rm Pl}^2}}, \label{eq:Hubble_parameter}
\eeq
with $a$ the scale factor, $m_{\rm Pl}=1.2 \times 10^{19}$ GeV the Planck mass and $\rho$ the total energy density of the universe.
The left hand side of the equations is a normalized rate of change of the corresponding comoving number density, $\frac{1}{a^3}\frac{d }{d t}(n a^3)$. The term proportional to $H$ simply captures the dilution due to the expansion of the universe.  
$n^{\rm eq}_N (T)$ is the would-be equilibrium number density of $N$ at a given temperature $T$.
$\langle\Gamma_N\rangle_T$ is the thermal average of the total decay rate of $N$ and $\epsilon$ is the CP-violating parameter that gives a measure of the (average) lepton number created per each decay of each $N$. These latter two model-dependent quantities will be computed in terms of the correlation functions of the strong sector in the next subsection (section~\ref{subsec:computeepsilonGamma}). 
Finally, $c$ is a constant characterizing the CFT, as discussed below.

Crucial to establishing Boltzmann equations \eqref{eq:BoltzmanneqNLV} and \eqref{eq:BoltzeqAsymLV} are three assumptions:
\begin{itemize}

    \item[1)] $\lambda$ must be ``small'', as quantified more precisely in section~\ref{sec:offshellwashout}. 
    
    Indeed, the coupling $\lambda$ is the spurion controlling lepton number as well as $N$-number violation: both $n_L$ and $n_N$ are conserved in the limit $\lambda \rightarrow0$. Hence, the key hypothesis that $n_L$ and $n_N$ are ``slowly varying densities'' is equivalent to the small $\lambda$ limit. Consistently, we note that in the Boltzmann equations we have dropped terms that are high order in $\lambda$. The most important correction among the neglected contributions arises from $\Delta L=2$ processes, and describes a new washout term proportional to $n_L$ on the right hand side of eq.~\eqref{eq:BoltzeqAsymLV}. We will identify sufficient conditions for this effect to be negligible in section~\ref{sec:offshellwashout}. When those conditions are satisfied our Boltzmann equations accurately approximate the physics we are interested in.

    \item[2)] The CFT is thermalized by the sizable self-couplings of the strong sector, and can be characterized by a temperature $\sim T$ of the order of the thermal bath and a chemical potential $\mu$ for the lepton number. 
    
    Given that the overall asymmetry in our models will always stay small during the entire evolution, we expect the linear relation $n_L=c\mu T^2/6$ to hold, where $c$ can be thought of as the effective number of degrees of freedom carrying lepton number in the CFT. For our analysis we will assume $c$ to be $\mathcal{O}(N_c)$, which is a model-dependent quantity appearing in eq.~\eqref{eq:BoltzeqAsymLV}.

    \item[3)] The generation of the asymmetry takes place dominantly at temperatures $T$ much smaller than the lightest of the $N_i$, of mass $m_{N_1}$. 
    
    This will be ensured by restricting our attention to a particular region of parameter space and/or making hypotheses on the history of the universe, as discussed in subsections \ref{sec:strongwashout} and \ref{sec:weakwashout}. 
    Once our hypotheses are satisfied, an accurate estimate of the asymmetry can be obtained by solving the equations from $T\ll m_{N_1}$, irrespective of whether or not at earlier times the universe has reached $T \geq m_N$ or $N$ has ever been relativistic or not.

    The assumption that the relevant temperatures for which eqs.~\eqref{eq:BoltzmanneqNLV} and \eqref{eq:BoltzeqAsymLV} are solved satisfy $T\ll m_{N_1}$ has important practical consequences. In particular it allows us to treat $N$ as non-relativistic. In practice this means it will be enough to consider the simpler equations governing the evolution of the number density $n_N$, rather than the more involved ones describing the evolution of the full distribution function $f_N$.\footnote{For a free Majorana fermion $N$, the number of particles carrying a given comoving momentum is conserved and therefore there is correspondingly a  set of infinite slowly varying variables captured by $f_N(\vec{p})$ for small $\lambda$. However the nonrelativistic condition allows us to track only the total number of $N$ particles.} Another important simplification following from $T\ll m_{N_1}$ is that the thermal corrections to \eqref{eq:BoltzmanneqNLV} and \eqref{eq:BoltzeqAsymLV} can be shown to be suppressed by powers of $T/m_{N_1}$ in all our model, and therefore to be negligible in that regime, as explained in appendix~\ref{app:finiteT}. 

\end{itemize}

The form acquired by eqs.~\eqref{eq:BoltzmanneqNLV} and \eqref{eq:BoltzeqAsymLV} is justified in the appendix via two complementary approaches. In appendix \ref{app:nearequilibrium} we adopt the near-equilibrium formalism developed in \cite{Bodeker:2014hqa} and \cite{Bodeker:2017deo} for the standard Type-I seesaw leptogenesis. That formalism is sufficiently general to be applicable to the models we consider and allows us to express all rates in terms of correlation functions of the CFT charge operator and of $O$. These correlation functions should in general be evaluated at finite $T$; however thanks to assumption 3) above and the results of appendix \ref{app:finiteT}, the rates can approximately be obtained from the zero temperature correlators. This way one finds that the Boltzmann equations reduce to those shown in eqs.~\eqref{eq:BoltzmanneqNLV} and \eqref{eq:BoltzeqAsymLV}, as claimed. Another derivation of the Boltzmann equations is offered in appendix \ref{app:moreintuitivederivation}. 
This second approach is admittedly not as systematic and rigorous as the first, but has the merit of shedding light and clarifying intuitively the underlying physics as well as the assumptions behind our analysis.

\subsection{Suppression of  $\Delta L=2$  washout processes }
\label{sec:offshellwashout}
The leading order interactions of decay and inverse decays of $N$ are considered in eqs.~\eqref{eq:BoltzmanneqNLV} and \eqref{eq:BoltzeqAsymLV} and will be used for the analysis in section~\ref{sec:evolution_asymmetry_high_scale}.
Let us now quantify the condition for small coupling mentioned earlier, that allows us to ignore higher order corrections in $\lambda$ to these equations. 

We focus on the washout effects mediated by ``off-shell'' $N$ at temperatures below $m_N$, which are the most relevant ones neglected by our equations. Such effects can be captured by the operator $\frac{\lambda^2}{m_{N}}O^2$, which violates lepton number by two units. 
At RG scales $\mu$ below $m_N$, we can define the dimensionless coupling associated with $O^2$ as $\bar{\kappa}_{_{\Delta L=2}}(\mu)$ given by 
\beq
\bar{\kappa}_{_{\Delta L=2}}(\mu)\equiv \bar{\lambda}^2(m_N) \left( \frac{\mu}{m_N} \right)^{2 \Delta-4},
\eeq
where, as done in section \ref{sec:neutrinomass}, we assumed a large-$N_c$ counting within the CFT, hence used that the scaling dimension of $O^2$ is approximately $2 \Delta$.
Below the confinement scale, $m_\rho$, lepton number violation is captured by the Weinberg operator whose coefficient is fixed by the neutrino mass. Therefore matching the two at the confinement scale, we can fix $\bar{\kappa}$, up to the elementary-composite mixing  of the neutrinos (see eq.~(\ref{eq:m nu gtr 25})):
\beq
m_\nu \sim s_{\nu}^2 \, \bar{\kappa}_{_{\Delta L=2}}(m_\rho) \frac{v^2}{m_\rho}.
\eeq
We can now use this result to estimate the rate of the washout process controlled by $O^2$ at temperatures $m_\rho\ll T\ll m_N$:
\beq
\Gamma^{\rm wo}_{\Delta L=2}(T)\sim [\bar{\kappa}_{_{\Delta L=2}}(T)]^2\, T\sim \frac{m_\nu^2 m_\rho^3}{s_{\nu}^4 v^4}  \left( \frac{T}{m_\rho} \right)^{4 \Delta-7}.
\eeq
Note that in particular for $\Delta=5/2$, corresponding to the  dimension of  $ O^2$ being $5$, just like the Weinberg operator, the $T$-dependence of the rate is the same as that in the standard Type-I seesaw. However, the dependence on $T$ may differ significantly for $\Delta\neq5/2$. Moreover, the rate is always enhanced parametrically compared to the standard seesaw scenario by $1/s_{\nu}^4$.

As well known, it is the ratio of the washout rate to the Hubble rate $H$
that determines whether the washout processes are in equilibrium. 
Since these washout processes are relevant when 
radiation dominates the cosmic energy density in eq.~\eqref{eq:Hubble_parameter}, we take
\beq
H = \sqrt{\frac{4\pi^3 g_\star}{45}} \frac{T^2}{m_{\rm Pl}},\label{eq:Hubble_parameter_radiation}
\eeq
where $g_*$ is the number of relativistic degrees of freedom.\footnote{For the CFT the effective number of degrees of freedom can be read from the energy momentum tensor.}
We identify two different regimes depending on whether $\Delta>9/4$ or $\Delta<9/4$: 
\begin{itemize}
\item[(i)] For $\Delta<9/4$, ${\Gamma^{\rm wo}_{ \Delta L=2}}/{H}$ increases as $T$ drops and its effect is dominated by $T$ close to $m_\rho$. Therefore, the constraint imposed by the requirement that these washout processes be out of equilibrium, and therefore irrelevant, reads
\bea
\frac{\Gamma^{\rm wo}_{ \Delta L=2}(m_\rho) }{H(m_\rho) } \ll 1 \implies s_\nu\gg s_{\nu,{\text{min}}}&\equiv& \left(\frac{45}{4\pi^3 g_\star}\right)^{1/8}
\left(\frac{m_\nu^2 m_{\rm Pl} m_\rho}{v^4}\right)^{1/4} \nonumber \\
&\sim &
10^{-2}  \left(\frac{100}{g_\star}\right)^{1/8}
\left(\frac{m_\nu}{0.05\,{\rm eV}}\right)^{1/2}
\left( \frac{m_\rho}{10 \, {\rm TeV}} \right)^{1/4}.
\label{eq:bound_offshell_wo_1}
\eea
\item[(ii)] For $\Delta>9/4$, the rate drops faster than the Hubble rate as $T$ decreases and the effect of the washout is most effective at high scale, near $T\sim m_N$. For this case a larger $s_\nu$ and a smaller $m_N$ are necessary to guarantee that $\Delta L=2$ processes be negligible: 
\beq\label{eq:bound_offshell_wo_2}
\frac{\Gamma^{\rm wo}_{\Delta L=2}(m_N) }{H(m_N) } \ll 1 
\implies s_\nu\gg s_{\nu,{\text{min}}}\times \left(\frac{m_N}{m_\rho}\right)^{\Delta-9/4}.
\eeq
where now the lower bound on $s_\nu$ is enhanced by a factor of $\left({m_N}/{m_\rho}\right)^{\Delta-9/4}$ compared to the case $\Delta<9/4$. \footnote{ The condition \eqref{eq:bound_offshell_wo_2} may also be expressed as an upper bound on the scaling dimension $\Delta$, namely
\bea
 \left( \Delta- \frac{5}{2} \right)  < 0.07 \frac{1+\log{s_{\nu}}-\frac{1}{4}\log{\frac{m_N}{10^{10}\, {\rm GeV}}} }{1+0.07\log{\frac{m_N/(10^{10}\, {\rm GeV})}{m_\rho/(10 {\rm TeV})}}}.
 \eea}
\end{itemize}

The estimates in eqs.~\eqref{eq:bound_offshell_wo_1} and \eqref{eq:bound_offshell_wo_2} capture the essential and robust parametric dependence of the washout rates but they ignore possibly large numerical factors. Therefore we will take them as qualitative indications of the allowed parameter space. The constraint from $\Delta L=2$ washout is shown as the grey region in the $m_N {\rm -} \Delta$  plane in figure~\ref{fig:Highscale_asymmetry} for two representative choices of $s_{\nu}$. For our numerical analysis we will choose benchmark points far from the boundary of the grey region in figure~\ref{fig:Highscale_asymmetry}. In those cases these washout processes are expected to be out of equilibrium and have negligible effects on the asymmetry.

To summarize, the irrelevance of $\Delta L=2$ washout effects results in important constraints on the parameter space of our models. First, the coupling of $N$ to the strongly coupled sector should be relevant or nearly marginal. In fact, for $\Delta$ much larger than $9/4$ high scale leptogenesis becomes progressively more constrained.
Second, the partial compositeness of the SM neutrinos can not be too small. More concretely we see that $s_\nu\gtrsim 0.01$ is favored.

\subsection{Computation of $\Gamma_N$ and $\epsilon$} 
\label{subsec:computeepsilonGamma}

In the previous subsection we obtained the Boltzmann equations for evolution of lepton number and the number density of $N$ in terms of two quantities denoted as $\Gamma_N$ and $\epsilon$. We will next compute these quantities.

To proceed we rewrite the model described by eq.~\eqref{eq:B-L_breaking} as follows
\begin{eqnarray}\label{LagDirac}
{\cal L} & \supset 
& \frac{1}{2}\overline{\Psi}_{N_i} (i \gamma^\mu \partial_\mu -m_{N_i})\Psi_N -\lambda_{i \alpha} \overline{\Psi}_{N_i} P_LO_\alpha -\lambda^*_{i \alpha} \overline{O}_\alpha P_R\Psi_{N_i},
\end{eqnarray}
where we have used 4-component spinor
\beq
\Psi_{N_i}= 
\begin{pmatrix}
N_i \\
i \sigma_2 N_i^*
\end{pmatrix},
\eeq
which satisfies $\Psi_N^c= C\overline{\Psi}_N^t= \Psi_N$, with $C$ being the charge conjugation matrix, $C=i\gamma^0 \gamma^2$. 
Without loss of generality we are working in the basis where the mass matrix for $N$ is diagonal with positive eigenvalues which we have denoted as $m_{N_i}$. Also $O_{\alpha}$
are left-handed fermionic SM-singlet CFT operators for which we have chosen a basis where they have well defined scaling dimension
$\Delta_\alpha$.
In the subsequent computations we are going to need the two-point correlators for $\Psi_N$ and $O$. The Feynman propagator for $\Psi$ is 
\beq
\langle\Psi_N\overline{\Psi_N}\rangle (p) = i\frac{\slashed{p}+m_N}{p^2-m_N^2+i\varepsilon}
\eeq
and the Fourier transform of the time-ordered two point correlator for the CFT operator $O$ is given by  
\bea\label{propagator}
\langle O_\alpha\overline{O}_\beta\rangle (p) &=& \delta_{\alpha\beta}iP_L\slashed{p}\frac{\mathcal{C}_\alpha}{\sin\pi(\Delta_\alpha-5/2)}(-p^2-i\epsilon)^{\Delta_\alpha-5/2}+{\cal O}\left[\slashed{p}\Lambda^{2\Delta-5}\left(\frac{p^2}{\Lambda^2}\right)^n\right] \nonumber \\
&\equiv&-\delta_{\alpha\beta}iP_L\slashed{p}~D_\alpha(p^2),
\eea
where $n=0,1,2,\cdots$ is a positive integer and the factor $1/\sin\pi(\Delta_\alpha-5/2)$ is introduced for convenience, so that
\bea
\textrm{Im}(D_\alpha(p^2))={\cal C}_\alpha|p^2|^{\Delta_\alpha-5/2}.
\eea
The parameter ${\cal C}_\alpha$ is real and positive, as will be determined below. The remainder in eq.~\eqref{propagator} arises because the Fourier transform of $\slashed{x}/x^{2\Delta+1}$ is UV divergent when $\Delta>5/2$. The divergence is local, in the sense that it is proportional to positive powers of $p^2$. As usual, the latter may be removed by counterterms, but in our case this will not be necessary because polynomial terms do not have an imaginary part and will not contribute to our analysis. 

\subsubsection{The decay rate $\Gamma_N$}
\label{subsubsec:N decay rate}

We now have all the ingredients necessary to evaluate the quantity $\langle\Gamma_N\rangle_T$ appearing in the Boltzmann equations \eqref{eq:BoltzmanneqNLV} and \eqref{eq:BoltzeqAsymLV}. Physically, at leading order in the parameters $\lambda_f$ mixing the SM and the strong sector, that corresponds to the thermal average of the zero-temperature decay rate $\Gamma_N$ of $N$ into the strong sector. The computation of the (zero-temperature) matrix element ${\cal M}(N_i\to {\text{Res}}_n)$ for $N_i$ decaying into resonances of the strong sector, labelled by the index $n$, is not obvious as one would need to know the overlap between $O$ and the resonances themselves, namely the wavefunction $\langle {\text{Res}}_n|\overline{O}_\alpha|0\rangle=\overline{u}_{\alpha,n}$. Fortunately, here we are interested in an inclusive rate, that sums over all possible final states of charge $+1$ and interpolated by $O$ (analogously, all those of charge $-1$ interpolated by $\overline{O}$). For that we can invoke the optical theorem and we can proceed without worrying about the explicit form of $u_{\alpha,n}$, as we will see shortly.

Consider $N$ decays into fermionic resonances ${\text{Res}}_n$ of charge $+1$ interpolated by $O$. The amplitude formally reads
\bea\label{amplitudeRes}
{\cal M}_{N_i\to CFT_n}=-i{\cal A}_{i\alpha}\,\overline{u_{n,\alpha}}P_Ru_N,
\eea
where ${\cal A}_{i\alpha}$ can in principle be calculated perturbatively for small coupling $\lambda_f$. The leading order contribution ${\cal A}_{i\alpha}=\lambda^*_{i\alpha}$ immediately follows from eq.~\eqref{LagDirac}. The next to leading corrections in $\lambda$ can be determined in terms of the $n$-point functions of the operator $O$. Under the assumption that $O$ has only 2-point functions, and ignoring corrections from all $\lambda_f$'s other than $\lambda$, we find 
\bea\label{asym}
{\cal A}_{\alpha i}&=&\lambda^*_{i \alpha}-\lambda^*_{j \alpha}\left[\frac{m_{N_i}^2}{m_{N_i}^2-m_{N_j}^2+i\varepsilon} \lambda_{j\beta} D_\beta \lambda^*_{i \beta }\right]
-\lambda^*_{j \alpha }\left[\frac{m_{N_i} \, m_{N_j}}{m_{N_i}^2-m_{N_j}^2+i\varepsilon} \lambda^*_{j\beta}D^t_\beta \lambda_{i \beta }\right]+{\cal O}(\lambda^5). \nonumber 
\eea
It is useful to compare this result to the one obtained in the familiar example of standard Type-I leptogenesis. In that language the operator $O$ can be identified with the composite SM-singlet fermionic operator $HL$. It is well-known that there are two  1-loop contributions to $N$ decays: the first one is from the vertex-correction diagram and the second is from the one-loop self-energy diagram. The first contribution can be seen to involve connected 4-point correlators of $O$, while the self-energy contribution involves only the two-point functions. In our large-$N_c$ counting we thus have only the latter contribution.

The inclusive rate $\Gamma(N\to O)$ is obtained by squaring \eqref{amplitudeRes} and integrating over the phase space of the ${\text{Res}}_n$ states, and finally summing over all $n$'s. 
Yet, the same result is related by the optical theorem to the imaginary part of the rate for $N_i\to N_i$ mediated by states of charge $+1$. Explicitly:
\beq
\Gamma(N_i\to O) = \frac{1}{m_{N_i}} \textrm{Im} \left.\mathcal{M} (N_i\rightarrow N_i)\right|_{\text{mediated by $O$}}.
\eeq
We see that to compute what we are interested in we do not need an explicit expression for $u_{\alpha, n}$ nor of the phase space of the resonances. Rather, the $n$-point functions of $O$ suffice. At leading order in $\lambda$ we just use the 2-point function of $O$ and get
\beq
\left.i\mathcal{M}(N_i\rightarrow N_i)\right|_{\text{mediated by $O$}}= i \lambda_{i \alpha} \lambda^*_{i \alpha} \overline{u} P_L \slashed{p} u D_\alpha(p),
\eeq
from which we immediately obtain the leading order expression
\begin{align}
\Gamma(N_i\to O)
&=\lambda_{i\alpha}\lambda^*_{i\alpha}\frac{m_{N_i}^2}{E}{\text{Im}}[D_\alpha(m_{N_i}^2)]\\\no
&=\lambda_{i \alpha}\lambda^*_{i \alpha}  \frac{m_{N_i}}{E}\mathcal{C}_{\alpha}m_{N_{i}}^{2\Delta_\alpha-5},
\end{align}
where a sum over the index $\alpha$ is left implicit. This equation reveals that a positive decay rate for $N_i$ requires $\mathcal{C}_\alpha>0$ (see ref.~\cite{Grinstein:2008qk} for a similar argument), as anticipated below eq.~\eqref{propagator}. In our large $N_c$ theory, the higher order corrections in $\lambda$ are encoded by simply replacing $\lambda_{i\alpha}^*$ with ${\cal A}_{i\alpha}$. One thus obtains the more general expression
\bea
\Gamma(N_i\to O)={\cal A}_{i\alpha}{\cal A}_{i\alpha}^*\frac{m_{N_i}^2}{E}{\text{Im}}[D_\alpha(m_{N_i}^2)].
\eea
Analogously, we can compute the partial width of $N_i$ into resonances of charge $-1$. The result reads 
\bea
\Gamma(N_i\to \overline{O})=\overline{\cal A}_{i\alpha}\overline{\cal A}_{i\alpha}^*\frac{m_{N_i}^2}{E}{\text{Im}}[D_\alpha(m_{N_i}^2)],
\eea
where in the basis where $m_{N_i}$ is real and diagonal and $D_\alpha$ is diagonal, that we use throughout the paper, $\overline{\cal A}_{i\alpha}$ is the same as ${\cal A}_{i\alpha}$ provided one replaces $\lambda$ with $\lambda^*$. The total width at zero temperature is hence given by
\beq
\Gamma_{N_i}=\Gamma(N_i\to O)+\Gamma(N_i\to\overline{O}),
\eeq

The thermal average of this quantity is just the decay width in the proper frame of $N_i$ times the thermal average of the Lorentz factor 
\beq
\left\langle \frac{m_{N_i}}{E} \right\rangle_T = \frac{1}{n_N^{\rm eq}} \int \frac{d^3 p_{\rm \scriptscriptstyle N}}{(2\pi)^3} f_F(E_N) \frac{m_N}{E_N} = \frac{K_1 (x)}{K_2 (x)}, \;\;\; x = m_N / T
\eeq
where $K_{1,2}(x)$ are modified Bessel function of the second kind, and the last equality holds for $f_F(E) = e^{-E/T}$. In the non-relativistic limit $x \to \infty$ we consistently recover the zero temperature expression in the rest frame, so the expression $\langle\Gamma_N\rangle_T$ appearing in eqs.~\eqref{eq:BoltzmanneqNLV} and \eqref{eq:BoltzeqAsymLV} can be written as $\langle\Gamma_N\rangle_T=\langle\Gamma_N\rangle_0\langle m_{N_i}/{E} \rangle_T$.

Our expressions can be made a bit more compact if we introduce the following dimensionless coupling 
\beq \label{eq:dimlesscoupling}
\tilde{\lambda}_{i \alpha}=\sqrt{\mathcal{C}_\alpha}\lambda_{i \alpha} m_{N_i}^{\Delta_\alpha-\frac{5}{2}},
\eeq
which up to the factor $\sqrt{\mathcal{C}_\alpha}$ coincides with the coupling shown in eq.~\eqref{eq:lambda_bar} renormalized at $\mu=m_{N_i}$. With our new notation we can for example express the zero-temperature decay rate at leading order in $\lambda$ as
$\langle\Gamma_{N_i}\rangle_0= \tilde{\lambda}_{i \alpha}\tilde{\lambda}^*_{i \alpha}  m_{N_{i}}$. Naive considerations suggest that $\tilde\lambda^2\sim\frac{N_c}{16\pi}\bar\lambda^2$.\footnote{According to naive dimensional analysis the propagator should scale as $\langle O\overline{O}\rangle={\mathfrak{c}}\frac{N_c}{16\pi^2}P_L{\slashed{p}}[(-p^2)^{\Delta-5/2}+\cdots]$ with ${\mathfrak{c}}$ some number that in general depends on $\Delta$. Comparing to eq.~\eqref{propagator} we thus have $\frac{{\cal C}}{\sin\pi(\Delta-5/2)}={\mathfrak{c}}\frac{N_c}{16\pi^2}$, but the parameter ${\mathfrak{c}}$ is undetermined by naive considerations. As a concrete example, consider a free CFT with $N_c$ families of a scalar ${\cal H}$ and a fermion $L$ in spacetime dimensions $4+\varepsilon$, with $\varepsilon\ll1$, and evaluate the 2-point function of $O={\cal H} L$. In that case we have $\Delta=5/2+\varepsilon$ and $\langle O\overline{O}\rangle\sim\frac{N_c}{16\pi^2}\frac{1}{\varepsilon}P_L{\slashed{p}}[(-p^2)^{\varepsilon}+\cdots]$ so that ${\mathfrak{c}}\sim \frac{1}{\varepsilon}=\frac{\pi}{\sin\pi(\Delta-5/2)}$ and ${\cal C}\sim\frac{N_c}{16\pi}$. For simplicity we will use the same estimate in the rest of the paper even when $\Delta$ significantly departs from $5/2$.
}

\subsubsection{The CP violation parameter $\epsilon$}
\label{subsubsec:CPV epsilon}

A natural generalization of the familiar definition of the parameter $\epsilon_i$, associated to the CP violation in $N_i$ decays, is given by
\beq\label{epsilonidef}
\epsilon_i=\frac{\Gamma(N_i\to O)-\Gamma(N_i\to\overline{O})}{\Gamma(N_i\to O)+\Gamma(N_i\to\overline{O})}.
\eeq
The correctness of this definition is proven a posteriori using the results in appendix \ref{app:source}. 

The explicit value of $\epsilon_i$ can be easily derived employing the results of section \ref{subsubsec:N decay rate}. Up to ${\cal O}(\lambda^4)$, and explicitating the sums over the flavor indices for clarity, we get 
\begin{align}\label{eq:epsilonasymCFT}
\epsilon_i 
&=\frac{\frac{1}{2} m_{N_i} \sum_\alpha   \textrm{Im}(D_\alpha(p))|_{p^{2}=m_{N_{i}}^{2}} \left[|{\cal A}_{\alpha i}|^2-|\overline{\cal A}_{\alpha i}|^2 \right]}{ m_{N_i} \sum_\alpha \lambda_{i \alpha} \lambda^*_{i \alpha}  \textrm{Im}(D_\alpha(p))|_{p^{2}=m_{N_{i}}^{2}}}\\\no
&= \sum_{\alpha \beta , j\neq i}{\rm Im}[D_\alpha] {\rm Im}[D_\beta] \left[\frac{2m_{N_i}m_{N_j}}{m_{N_i}^2-m_{N_j}^2}\frac{ {\rm Im}[\lambda^*_{j\beta}\lambda_{i \beta}\lambda^*_{j\alpha}\lambda_{i \alpha} ]} {{(\tilde\lambda\tilde\lambda^\dagger)_{ii}}} +\frac{2m_{N_i}^2}{m_{N_i}^2-m_{N_j}^2} \frac{{\rm Im}[\lambda_{j\beta}\lambda^*_{i \beta}\lambda^*_{j\alpha}\lambda_{i \alpha} ]}{{(\tilde\lambda\tilde\lambda^\dagger)_{ii}}} \right] \nonumber \\
 &= \sum_{\alpha \beta, j\neq i}{\rm Im}[D_\alpha] {\rm Im}[D_\beta] \frac{2m_{N_i}m_{N_j}}{m_{N_i}^2-m_{N_j}^2} \frac{{\rm Im}[\lambda^*_{j\beta}\lambda_{i \beta}\lambda^*_{j\alpha}\lambda_{i \alpha} ]} {{(\tilde\lambda\tilde\lambda^\dagger)_{ii}}} \nonumber \\
 & =  \sum_{ j\neq i} \frac{2m_{N_i}m_{N_j}}{m_{N_i}^2-m_{N_j}^2} \sum_{\alpha \beta} \frac{{\rm Im}[\tilde\lambda^*_{j\beta}\tilde\lambda_{i \beta}\tilde\lambda^*_{j\alpha}\tilde\lambda_{i \alpha} ]} {{(\tilde\lambda\tilde\lambda^\dagger)_{ii}}} \left( \frac{m_{N_i}}{m_{N_j}} \right)^{\Delta_\alpha+\Delta_\beta-5}.
\end{align}
Note that the expression remains unchanged if we replace the partial widths in eq.~\eqref{epsilonidef} with their thermal averages.

Consistently with the argument given by Nanopoulos and Weinberg \cite{Nanopoulos:1979gx}, the CP-violating parameter $\epsilon_i$ starts at order $\lambda^2$. 
As a rough approximation for $\epsilon_i$, one may consider a scenario of nearly degenerate singlets, $m_{N_i}\sim m_{N_j}$, with anarchic and maximally CP violating couplings ${\text{Im}}[\lambda]\sim|\lambda|$. In that case we roughly have
\begin{eqnarray}\label{eq:epsilon_simplified}
    \epsilon\sim \tilde\lambda^2, 
\end{eqnarray}
We will often use this rough estimate for numerical estimates in our plots.

\section{ Generation and Evolution of the Asymmetry  at the High Scale}
\label{sec:evolution_asymmetry_high_scale}

In this section we are going to identify a number of scenarios in which the asymmetry is generated at temperatures $T\ll m_{N_1}$, so that the simplifying assumption 3) spelled out in section \ref{subsec:BELV} applies. For each of these scenarios we will then estimate the lepton asymmetry at temperatures $T\gg m_\rho$ above the phase transition.

In order to identify our scenarios it is useful to introduce a parameter conventionally known as the \emph{washout factor} $K$. In a radiation dominated universe with Hubble parameter given by eq.~\eqref{eq:Hubble_parameter_radiation}, the latter can be defined as 
\beq \label{eq:defineK}
K \equiv \left( \frac{\langle\Gamma_{N_1}\rangle_T}{H} \right)_{T=m_{N_1}} = \sqrt{\frac{45}{4 \pi^3}}\frac{\langle\Gamma_{N_1}\rangle_{T=m_{N_1}} m_{\rm Pl}}{\sqrt{g_*} m_{N_1}^2}\sim \sqrt{\frac{45}{4 \pi^3}} \frac{m_{\rm Pl}\tilde{\lambda}^2}{\sqrt{g_*} m_N},
\eeq
where 
in the last step we used the zero-temperature decay width of section \ref{subsubsec:N decay rate}.
Recalling that the neutrino mass is given by eq.~\eqref{eq:m nu gtr 25} we can rewrite $K$ in terms of the neutrino mass as
 \beq\label{eq:K in terms of m_nu}
    K\sim \left[ \sqrt{\frac{45}{4\pi^3 g_*}}\frac{ \pi}{g_\rho^2}\frac{m_\nu m_{\rm Pl}} {v^2} \right] \frac{1}{s_{\nu}^2} \left( \frac{m_N}{m_\rho}\right)^{2 \Delta-5} 
    \approx\left[20 \left(\frac{4 \pi}{g_\rho}\right)^2
    \left(\frac{m_\nu}{0.1\,\mathrm{eV}}
    \right)\sqrt{\frac{100}{g_*}}\,\right]
    \frac{1}{s_{\nu}^2} \left( \frac{m_N}{m_\rho}\right)^{2 \Delta-5},
    \eeq
where the factor in the bracket is the same as that of the Type-I seesaw model. As we will see, the definition \eqref{eq:K in terms of m_nu} will still be convenient even when encountering situations in which the universe was not radiation dominated. 

With our new notation, the size of the washout term in eq.~\eqref{eq:BoltzeqAsymLV} is parametrized by the dimensionless combination $(n_N^{\text{eq}}/T^3)K/c$. We will refer to the regime of $K/c \gg 1$, with the additional assumption about the reheating temperature stated above, as the \emph{strong washout regime}. The other extreme case $K/c \ll 1$ will instead characterize the \emph{weak washout regime}. In the following subsections we will see that in both the strong and weak washout limits we can easily identify scenarios that satisfy our hypothesis 3). We begin by first identifying scenarios in the strong washout regime (see section \ref{sec:strongwashout}) and then move to the weak washout regime (see section \ref{sec:weakwashout}).

\subsection{Strong Washout} \label{sec:strongwashout}

In the regime of strong washout $N$ couples strongly enough to the CFT at temperatures of the order of its mass so that $N$ is essentially kept close in thermal equilibrium with the CFT sector at $T\sim m_N$. In this situation any asymmetry, either generated by $N$ decays or initially present, is efficiently washed out so long as the equilibrium is maintained. This situation continues to hold until the universe reaches a sufficiently low temperature $T \ll m_N$ such that the processes CFT $\to N$ freeze out and the two systems fall out of equilibrium. In this regime one can thus obtain a good estimate of the generated asymmetry by solving the Boltzmann equations of section \ref{subsec:BELV} starting from $T\ll m_N$ and with equilibrium initial conditions. The insensitivity to initial condition and to the previous history makes this an especially predictive and attractive regime.

Let us be a bit more quantitative. We first identify the region of parameter space in which our models feature a strong washout regime. We go back to the expression for $K$ in \eqref{eq:K in terms of m_nu} and stress two notable differences compared to standard leptogenesis. 
The first difference is the dimension of the leading operator that violates lepton number at energy scales below $m_N$. This difference leads to a universal RG factor $(m_N / m_\rho)^{2\Delta -5}$. This results in a sizable enhancement of $K$ for any $\Delta>5/2$.
The second difference is a factor $1/s_\nu^2$ coming from the elementary-composite mixing $s_\nu$ appearing in the neutrino mass formula. This also leads a contribution larger than unity since $s_{\nu}<1$ by construction.

\begin{figure}
    \centering
\includegraphics[width=7cm]{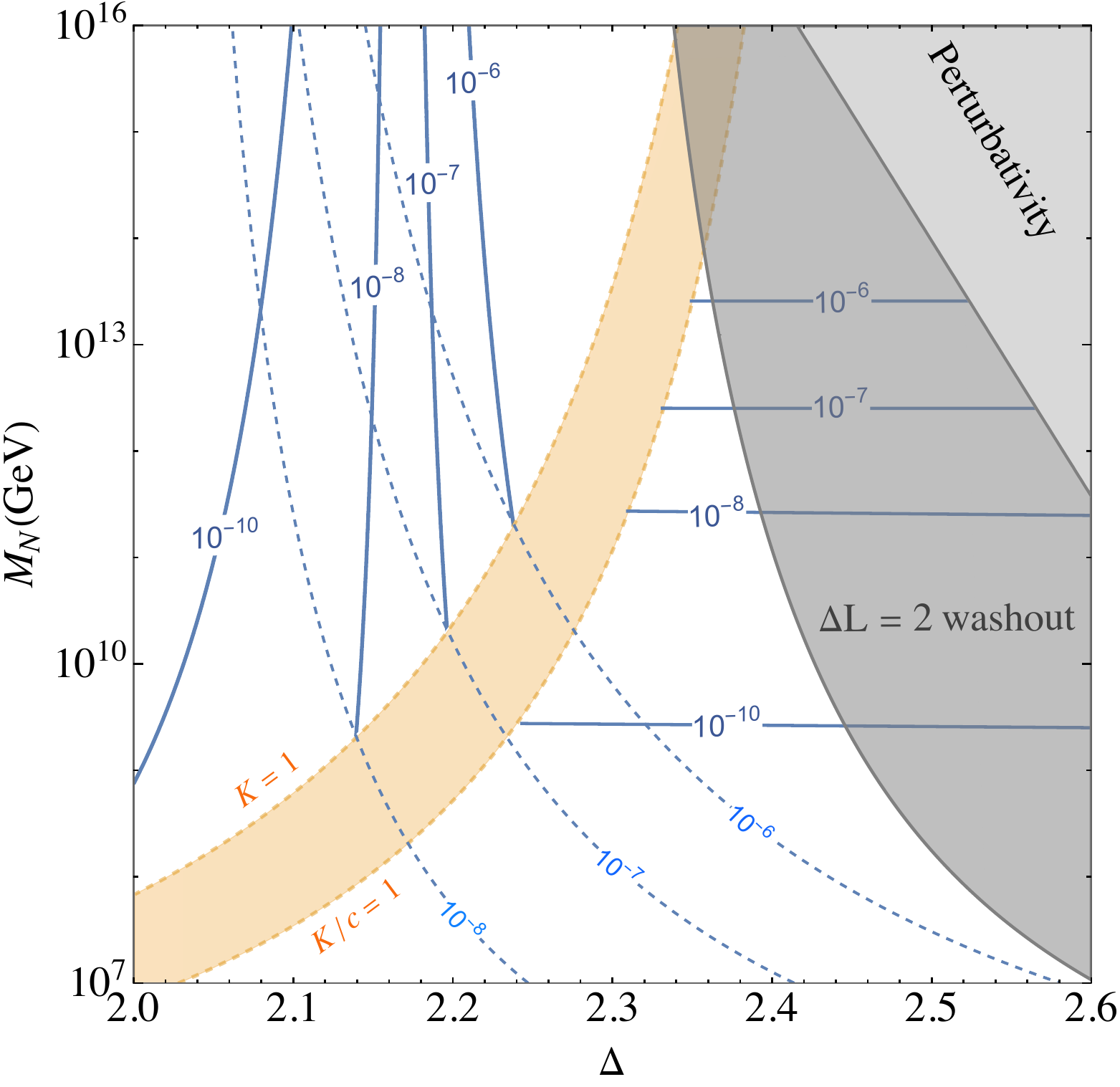}
\includegraphics[width=7cm]
{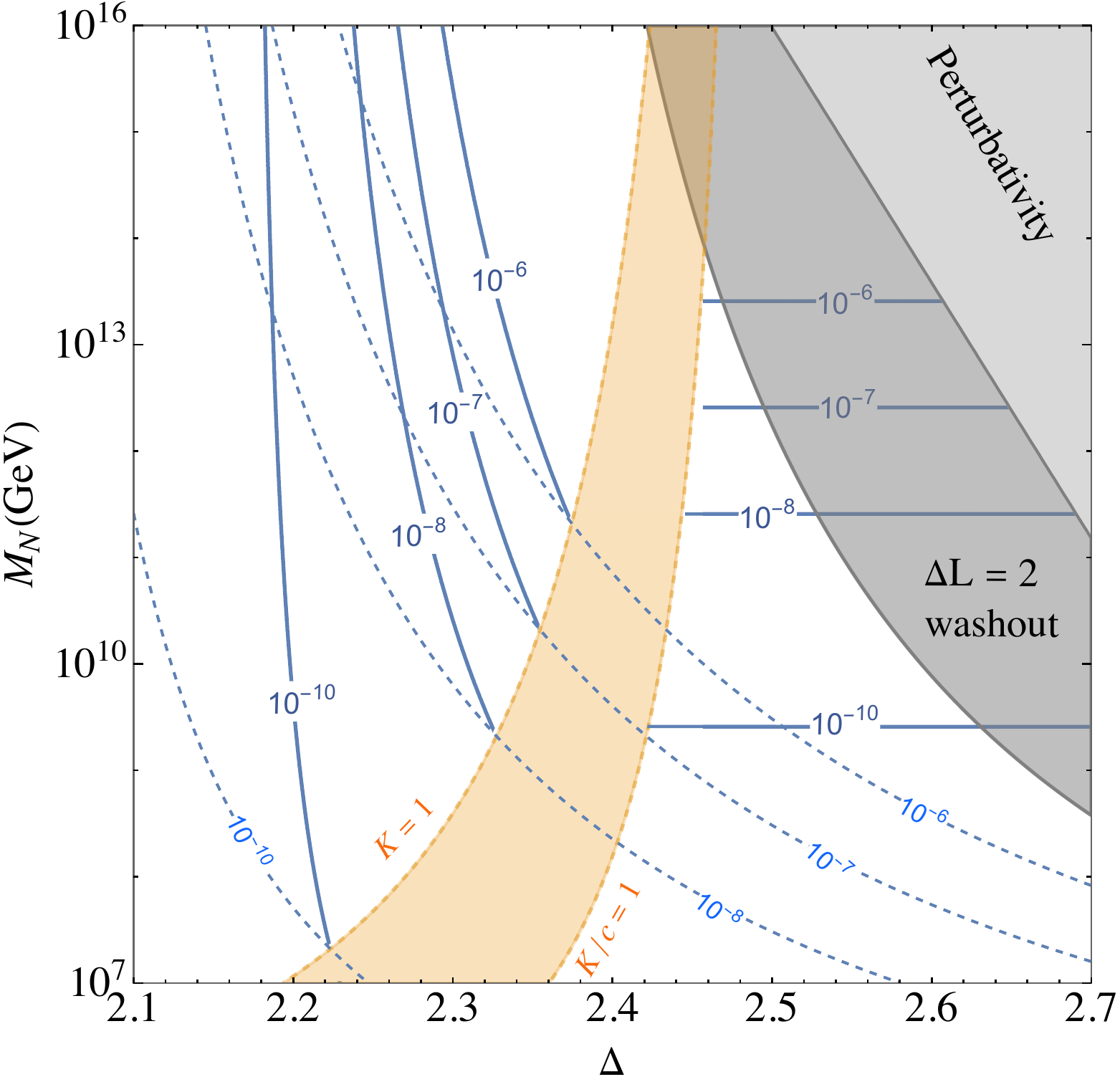}

    \caption{Allowed region in the $(m_N,\Delta)$ plane for high scale generation of the asymmetry. We set the coupling $\bar\lambda(m_N)$ in order to obtain a SM neutrino mass $m_\nu=0.05$ eV, we take $c=N_c=10$, $g_*=g_*^{\rm SM}+g_*^{\rm CFT}$~(see eq.~\eqref{eq:g_star_general}), and the partial compositeness of SM lepton doublet being $s_{\nu}=0.1$ (left) and $s_{\nu}=1$ (right). $K$ is the washout factor shown in eq.~\eqref{eq:defineK}. Solid blue contours denote the asymmetry generated from $N$ decays ($Y_{L}^{\rm UV}$). Dashed light blue contours show the value of the CP parameter $\epsilon$ assuming $\bar\lambda(m_N)$ is anarchic in flavor, see eq.~\eqref{eq:epsilon_simplified}. In the strong washout regime we used eq.~\eqref{eq:strong_wo_UV} whereas in the weak washout regime we used the representative value quoted in eq.~\eqref{eq:Y_L estimate weak wo}. In the  shaded orange region the washout factor is neither large nor small and a more refined analysis would be needed. 
    The shaded gray regions are excluded by the naive perturbative requirement $\bar{\lambda}(\Lambda)<4\pi/\sqrt{N_c}$ (light gray),
    and by the presence of too much washout from $\Delta L=2$  processes (dark gray).} 
    \label{fig:Highscale_asymmetry}
\end{figure}

The bottom line is that the regime $K/c\gg1$ can be realized with a relevant coupling $\lambda N O$ if $s_\nu\ll1$ and $m_N/m_\rho$ is not too large. The strong washout condition $K/c\gg1$ is instead generic whenever $\lambda N O$ is marginal or irrelevant. Yet, in the latter case the conditions of section \ref{sec:offshellwashout} necessary to neglect the higher order terms in our Boltzmann equations are non-trivial and significantly limit the parameter space. These observations are manifest in figure~\ref{fig:Highscale_asymmetry}.

Assume now $K/c \gg 1$ and consider the Boltzmann equations eqs.~\eqref{eq:BoltzmanneqNLV} and \eqref{eq:BoltzeqAsymLV}. From the first follows that $n_N$ remains close to the equilibrium distribution for temperatures well below $m_{N_1}$. The second equation reveals that the washout rate effectively depletes any asymmetry as long as $(n_N^{\text{eq}}/T^3)(K/c)\gtrsim1$, which using the approximation $n_N^{\text{eq}}/T^3\sim e^{-m_T/T}$ means until $T\sim m_{N_1}/\textrm{ln}(K/c)\ll m_{N_1}$. We conclude that a good description of the genesis may be obtained solving the Boltzmann equations for $T\ll m_{N_1}$. According to Appendix \ref{app:finiteT}, approximating the CFT correlators with their zero-temperature expressions is thus correct up to thermal corrections of $\mathcal{O}\left( 1/ \textrm{ln}^3 (K/c)  \right)$. The logarithmic dependence may at first sight seem to suggest a poor convergence of the expansion. However, already for $K/c\sim 10$ this correction is $\mathcal{O}(8 \%)$. 

Using the parameters $\epsilon$ and $\Gamma_N$ calculated in section~\ref{subsec:computeepsilonGamma} we can finally derive the asymmetry generated at high scale. The result can be simply obtained from that of the standard type-I seesaw leptogenesis by replacing $\epsilon$ and $K$ with those computed for our case. The asymmetry generated by the $N$ decay in the strong washout regime, which we refer it to as ``UV asymmetry'', is approximately given by \cite{Agashe:2018cuf,Buchmuller:2004nz} 
\beq\label{eq:strong_wo_UV}
Y_{L}^{\rm UV} \equiv \frac{n_L^{\rm UV}}{s} \sim \frac{c \, \epsilon}{g_* K} 
\eeq
where $s$ is the entropy density. Plugging eq.~\eqref{eq:epsilon_simplified} and eq.~\eqref{eq:defineK} into eq.~\eqref{eq:strong_wo_UV}, we obtain 
\begin{eqnarray}\label{eq:Y_DeltaL_UV_stong_wo}
    Y_{L}^{\rm UV}\sim \sqrt{\frac{4 \pi^3}{45}} \frac{c\, m_N}{\sqrt{g_*}m_{\rm Pl}}.
\end{eqnarray}
We show contours of constant $Y_{L}^{\rm UV}$ in the $(m_N,\Delta)$ plane of figure~\ref{fig:Highscale_asymmetry}. The contours to the right of the orange shaded region are well inside the strong washout regime ($K/c>10$). In this case eq.~\eqref{eq:Y_DeltaL_UV_stong_wo} accurately approximates $Y_{L}^{\rm UV}$ and all contours are independent of $\Delta$.

\subsection{Weak washout}  \label{sec:weakwashout}

The weak washout regime is defined by $K/c\ll1$. Recalling that $c={\cal O}(N_c)$ and that we always consider the large $N_c$ limit, it will be sufficient to take $K\lesssim1$.  

In this regime $N$ couples weakly to the CFT such that the coupling $\lambda$ cannot bring $N$ and the CFT to thermal equilibrium at temperatures of order $T\sim m_N$. To guarantee that the asymmetry is generated dominantly by $N$ decays at $T\ll m_{N_1}$ one has to make assumptions about the reheating process, and in particular on the initial lepton asymmetry as well as the initial abundances of $N$ and radiation (namely the CFT and the SM). In this work, we assume that no primordial asymmetry is present after inflation. If this is not the case, any pre-existing asymmetry may be easily accounted for by simply adding it to the contribution generated by our mechanism. Regarding the initial abundances of $N$ and radiation, we will next discuss three scenarios that guarantee that the hypothesis 3) of section \ref{subsec:BELV} be satisfied. Other options are clearly possible, though.

\subsubsection{Thermal $N$ in radiation-domination} 
In this first scenario the CFT and/or the SM are populated after inflation and the universe is assumed to be radiation-dominated. $N$ is brought to thermal equilibrium at approximately the same temperature as the bath by other (heavy) states. As $T$ drops below the mass of the heavy states, $N$ falls out of equilibrium while keeping a thermal distribution function until its decay.\footnote{Note that by thermal here we don't mean that the distribution function is that of equilibrium, but that  the distribution function of $N$ in terms of the comoving momenta is simply preserved from the moment that $N$ falls out of equilibrium until the decays become relevant.}  In this setup we are guaranteed that the asymmetry is dominantly induced by $N$ decays.

In radiation-domination the expansion rate can be written as $H(T)\sim(T/m_N)^2H(m_N)$ and the temperature at which the decay rate $\Gamma_N$ starts to be relevant, $H(T_{\text{dec}})\sim\Gamma_N$,  is  \beq\label{tdec}
T_{\text{dec}}\sim\sqrt{K} m_N,
\eeq
which is parametrically smaller than $m_N$ as long as $K\ll1$, as promised. To be self-consistent we have to verify that the decays take place in the radiation dominated era. This condition reads $g_*T_{\text{dec}}^4\gtrsim m_NT_{\text{dec}}^3$ and implies
\beq\label{raddomcond}
\frac{1}{g_*^2}\lesssim K\lesssim1.
\eeq
With $K$ in this range we can determine the asymmetry using the Boltzmann equations in section \ref{subsec:BELV} with an initial thermal population for $N$. Otherwise, for $K \lesssim1/g_*^2$, the universe would enter an era dominated by non-relativistic $N$ matter, a case that we consider subsequently.

The asymmetry in this scenario can be straightforwardly obtained from that of the standard leptogenesis in the same regime provided we use the CPV parameter $\epsilon$ obtained in \eqref{eq:epsilonasymCFT}. This way we get 
\beq
Y_{L}^{\rm UV} \sim \frac{\epsilon}{g_*}\gtrsim \epsilon \sqrt{K}.
\eeq
The result has a simple interpretation: a thermal abundance for $N$ leads to a yield of order $Y_N \sim 1/g_{*}$ at $T\gg m_N$. As the temperature decreases $Y_N$ is unchanged until its decay rate becomes comparable to $H$, namely when $T_{\text{dec}}^2\sim \Gamma_N m_{\text{Pl}}/\sqrt{g_*}$. From that moment on, an $\epsilon$ fraction of the $N$ abundance is converted to the asymmetry.

\subsubsection{Non-thermal $N$ in radiation-domination} 
    
In the second scenario the CFT and/or the SM are populated after inflation and the universe is radiation-dominated, as before. Yet, here $N$ is produced non-relativistically with a yield $Y_{N,\text{init}}$ larger than the thermal value.\footnote{If the yield is smaller than the equilibrium value the asymmetry generated by the production cannot be ignored, and a more dedicated study would be needed. When this condition is met in the type-I seesaw leptogenesis the contributions to the asymmetry from decay and production are known to cancel each other to leading order in $K$, leading to a final asymmetry $\mathcal{O}(\epsilon K^2)$. This may presumably happen in our scenario as well, but a confirmation would require the computation of the source terms for $T> m_N$, which seems technically challenging at the moment. We therefore decide to stay away from this regime.}

The requirement (analogous to the lower bound on $K$ in \eqref{raddomcond}) that the universe is radiation-dominated for all relevant temperatures is $m_N sY_N(T)\sim m_Ng_*Y_N(T)\lesssim g_*T^4$. The strongest condition arises at the temperature \eqref{tdec} when decays start, and reads $m_NY_{N,\text{init}}\lesssim T_{\text{dec}}$, or
\beq
Y_{N,\text{init}}\lesssim\sqrt{K}.
\eeq
When these assumptions are satisfied we can again use the Boltzmann equations in section \ref{subsec:BELV}. As a result the asymmetry is given by 
\beq
 Y_{L}^{\rm UV} \sim \epsilon Y_{N, \rm{init}}\lesssim\epsilon\sqrt{K}.
\eeq

 \subsubsection{Non-relativistic $N$ in matter-domination}

 In this third scenario $N$ is produced non-relativistically and its initial abundance dominates the universe. The CFT and other elementary particles are created as $N$ starts decaying. 
 
 Note that this setup encompasses a variety of possible initial conditions. Indeed, whatever the initial distribution for $N$ is, the expansion of the universe would redshift the momenta and eventually $N$ would acquire a non-relativistic distribution, unless $N$ decays earlier.\footnote{Ref.~\cite{Buchmuller:2011mw} considered a scenario where $N$ particles are produced from decays of heavy scalars after inflation, taking into account the transition from relativistic to non-relativistic $N$. Even in this scenario where $N$ had an initially relativistic distribution, they find that $N$ particles only decay when they become non-relativistic, due to time dilation with respect to the thermal bath frame.} In fact, our key assumption is that $N$ has a dominantly non-relativistic distribution at temperatures before $\Gamma_N$ becomes comparable to $H$.

 In this setup the only relevant process for genesis is $N$ decays, as no other particle is around, and by assumption these occur when $N$ is non-relativistic. The Boltzmann equations in section \ref{subsec:BELV} must be replaced by analogous ones in which the equilibrium densities are essentially neglected, plus of course a relation defining $H$ and the radiation:
   \bea\label{eq:mattdom}
   && \frac{d n_N}{dt}+ 3 H n_N \approx -  \Gamma_N n_N, \\
   && \frac{d n_L}{dt}+ 3 H n_L \approx  \epsilon \Gamma_N n_N, \\
   && H^2= \frac{8 \pi G}{3} \left( m_N n_N + \rho_{\rm rad}\right), \\
   && \frac{d \rho_{\rm rad}}{dt}+ 4 H \rho_{\rm rad} = \Gamma_N m_N n_N.
    \eea 
Let us first qualitatively analyze the implications of these equations under the assumption that the initial abundance of $N$ is sufficiently large, as quantified shortly. Then it turns out that the decays start to become relevant at a time $t_{\text{dec}}$ when the Hubble scale becomes comparable to the decay width. From that time on, $N$s decays produce an asymmetry and a thermal bath for the CFT and the SM states. The number density of $N$ at $t_{\text{dec}}$ can be estimated by the condition $\Gamma_N^2 \sim H^2_{\text{dec}} = \frac{8\pi G}{3} \rho_{N,\text{dec}}$ and reads
    \beq\label{eq:critical n_N dec}
    n_{N,\text{dec}}\sim\frac{3 \Gamma_N^2}{8 \pi G m_N}\sim \frac{\Gamma_N^2 m_{\rm{Pl}}^2}{m_N}. 
    \eeq
As a consistency check of our analysis, we find that the temperature of the thermal bath generated by the decay is of order $m_N n_{N,\rm{dec}} \sim g_* T_{\rm dec}^4$, that is
\beq
T_{\text{dec}}\sim\frac{\sqrt{\Gamma_N m_{\rm Pl}}}{g_*^{1/4}}.
\eeq
From eq.~\eqref{eq:defineK} follows that $T_{\rm dec}\ll m_N$ for $K\ll 1$, so we can confidently use our non-relativistic approximation.

The lepton number density generated in this scenario is intuitively given by $n_L \sim \epsilon n_{N, \rm{dec}}$, so the yield for the lepton asymmetry ($Y_{L}^{\rm UV}$) can be written as
    \beq\label{eq:Y_L estimate weak wo}
    Y_{L}^{\rm UV} \sim \frac{n_L}{g_* T_{\text{dec}}^3} \sim \epsilon \frac{\sqrt{\Gamma_N m_{\rm Pl}} }{g_*^{1/4} m_N}\sim\epsilon\sqrt{K}.
    \eeq
This estimate is confirmed by a more accurate numerical analysis of \eqref{eq:mattdom}. Importantly, 
our final result \eqref{eq:Y_L estimate weak wo} does not depend on the initial condition so long as the universe was dominated by the energy density of non-relativistic $N$'s prior to $N$ decays, and provided the initial number density of $N$ exceeded the critical value $n_{N,{\rm dec}}$ shown in \eqref{eq:critical n_N dec}. In that case \eqref{eq:Y_L estimate weak wo} is a robust result and reveals that the final asymmetry is completely fixed by $\epsilon$ and $K$. Hence, while in general the asymmetry in the weak washout regime is a priori strongly sensitive to the initial conditions, there are reasonable assumptions for which such sensitivity can be rather mild.

If $N$ still dominates the energy density but its initial abundance is smaller than the critical value, that is if $n_{N,\text{in}}\sim H^2m_{\text{Pl}}^2/m_N<n_{N,{\rm dec}}$, then we are by construction in a regime in which $\Gamma_N>H$ always. Hence $N$ decays right after reheating and the UV asymmetry is given by $n_L\sim\epsilon n_{N,\text{init}}$. Now the yield at decay is approximately $Y_{N}^{\rm UV}\sim {n_{N,\text{init}}^{1/4} }/({g_*^{1/4} m_N^{3/4}})$, and is thus reduced compared to the previous case by a factor $\left( n_{N,{\rm init}} / n_{N,{\rm dec}}\right)^{1/4}$. Therefore the final estimate for the UV asymmetry generated from $N$ decays is also smaller
\beq
Y_{L}^{\rm UV} \sim \left( \frac{n_{N,{\rm init}}}{n_{N,{\rm dec}}}\right)^{1/4} \epsilon  \sqrt{K}\lesssim \epsilon  \sqrt{K}.
\eeq
From eqs.~\eqref{eq:dimlesscoupling}, \eqref{eq:epsilon_simplified} and \eqref{eq:defineK}, we see that in this regime, $Y_L^{\rm UV} \propto m_N^{3\Delta - 13/2}$. In particular, note that in the limit  $\Delta = 13/6$, it becomes nearly independent of $m_N$, a feature we can also observe in figure~\ref{fig:Highscale_asymmetry}.

\section{Low Scale Dynamics}
\label{sec:low_scale_dynamics}

In previous section we have discussed and analyzed the generation and evolution of the asymmetry before the confinement phase transition during which the strongly coupled sector develops a mass gap $m_\rho$. In this section we discuss the impact of the phase transition as well as of the resonances formed after the PT. We find that both tend to suppress the present-day asymmetry compared to the value obtained in section \ref{sec:evolution_asymmetry_high_scale}. The suppression due to the reheating produced at the PT is parametrized by a factor $\eta_{\text{PT}}$ obtained in section \ref{sec:phase_transition}, whereas the suppression due to the low-energy washout induced by the resonances is parametrized by a factor $\eta_{\text{IR}}$ estimated in section \ref{sec:low scale washout}.

\subsection{Effects of the Confinement Phase Transition}
\label{sec:phase_transition}
We now discuss the features of confinement phase transition and its impact on the asymmetry. 
The confinement of the composite sector is associated with the breaking of conformal invariance, which can be captured by a deformation to the CFT that is dual to the Goldberger-Wise mechanism in the holographic 5D models~\cite{Goldberger:1999uk} (for details of CFT dual description of spontaneous breaking of conformal invariance and dilaton effective theory, see e.g. \cite{ArkaniHamed:2000ds, Rattazzi:2000hs, Agashe:2016rle}),
\beq \label{eq:GWdeformation}
\Delta \mathcal{L}_{\rm CFT}=g_{_{\rm GW}} \mathcal O_{\rm GW},
\eeq
 where $\mathcal O_{\rm GW}$ is a nearly-marginal CFT operator with scaling dimension $[\mathcal O_{\rm GW}]=4+\epsilon_d$.
 \footnote{In earlier sections we have also seen a deformation of the CFT generated after integrating out $N$ [see eq. \eqref{eq:CFTatmrho}], but that is generically an irrelevant operator with a larger dimension than $\mathcal O_{\rm GW}$ in order to produce small neutrino masses. Therefore its effect on scale-invariance breaking is subdominant and thus neglected in this section.} Generating a large hierarchy between the UV scale and the confinement scale, necessary to address the hierarchy problem and generate the flavor hierarchies, is achieved by a small $\epsilon_d$~\cite{Goldberger:1999uk,Rattazzi:2000hs}, 
\beq \label{eq:hierarchyepsilon}
\ln \frac{\Lambda}{m_\rho}  \sim 1/\epsilon_d
\eeq
where $m_\rho$ characterizes the typical mass scale of the composite states and we will simply refer to it as the confinement scale.

Now let us describe the cosmological history and the PT as the universe cools down (for early references, see for example~\cite{Creminelli:2001th, Randall:2006py, Nardini:2007me, Konstandin:2010cd, Konstandin:2011dr}). 
At high temperatures $T > m_\rho$, the strongly coupled sector is in a deconfined phase, 
while at sufficiently low temperatures $T \ll m_\rho$, the strongly coupled sector confines and produces composite ``hadrons''. 
Then, it is expected that as temperature cools below a critical temperature $T_c$, a transition from a deconfined phase to a confined phase occurs. The analysis of this PT for general CH theories is not possible. However, we can be rather concrete in the case of theories that admit an AdS/CFT dual description in terms of 5D Randall-Sundrum models.
The properties obtained from the holographic 5D picture can be understood in a rather robust way also from the 4D picture in models where the breaking of scale invariance is predominantly spontaneous \cite{Agashe:2019lhy}.
Given that no robust and controlled analysis is known so far for cases beyond these models, we will restrict our analysis to this class. 

Typically, this PT is first-order as two distinct phases can occur at the same temperature. Therefore, the PT can proceed by nucleation and percolation of bubbles. Although the nucleation of bubbles starts as $T$ drops below $T_c$, the PT completes only when bubbles percolate at the so-called nucleation temperature $T_n$ where the bubble nucleation rate is comparable to the Hubble rate. %
After the transition, the vacuum energy of the false vacuum will be converted into energies of particles in the composite as well as the elementary sectors. 
Therefore, SM bath will be heated up to a temperature $T_r$. During this reheating process, an entropy of $\mathcal{O}(T_r^3)$ is injected into the SM thermal bath and thus the yield of the asymmetry is diluted by a factor of the order $\sim (T_n/T_r)^3$.  This is the main effect of the PT on the lepton asymmetry generated from leptogenesis. In the rest of the section, we will discuss the determination of $T_c$, $T_n$ and $T_r$, and the size of the dilution factor in these models.

{ \bf The critical temperature} $T_c$ can be estimated from the equality of the free energy densities of the two phases. 
 As discussed in \cite{Agashe:2019lhy,Agashe:2020lfz}, the free energy density of CFT in the deconfined phase is given as
 \begin{eqnarray}\label{eq:F_T_deconf}
     F_{\rm deconfined}(T)=-\frac{\pi^2}{8} N_c^2 T^4,
 \end{eqnarray}
 where $N_c$ corresponds to the number of colors of the $SU(N_c)$ gauge group of CFT. The prefactor $\pi^2/8$ is model-dependent, and we take the value from its holographic dual 5D models~\cite{Agashe:2020lfz}.
$F_{\rm deconfined}(T)$ is basically the free energy density of a thermal radiation bath of $O(N_c^2)$ degrees of freedom.

In the confined phase at low temperature, the free energy is dominated by the potential energy of the lightest hadronic states.  These states include the dilaton, the pNGB of the spontaneous conformal symmetry breaking as well as possibly other pNGBs, corresponding to the spontaneous breaking of the global symmetries of the strongly coupled sector, including the composite Goldstone Higgs particles. We work in the regime where the dynamics of the confined phase at the phase transition is dominated by that of the dilaton.
The deformation of eq.~\eqref{eq:GWdeformation} leads to generation of a non-scale-invariant potential for  the dilaton. This is the AdS/CFT dual picture \cite{ArkaniHamed:2000ds, Rattazzi:2000hs, Agashe:2016rle} of the Goldberger-Wise stabilization mechanism in 5D warped extra dimension.
The Lagrangian of the dilaton field ($\phi$) in this case  has the form:
\beq \label{eq:dilatonpotential}
\mathcal L(\phi)=\frac{3 N_c^2}{4\pi^2}\left[\frac{1}{2}(\partial \phi)^2-\lambda_d \, \phi^4\left(1-\frac{1}{1+\epsilon_d/4}\left(\frac{\phi}{\langle\phi\rangle}\right)^{\epsilon_d}\right)\right],
\eeq
where $\langle\phi\rangle$ is the VEV of dilaton field at the minimum of the potential and we have chosen a (non-canonical) normalization for $\phi$ such that the typical mass of the composite states is $m_\rho \sim  \langle\phi\rangle$, without any additional $N_c$ or coupling dependence. 
 $\lambda_{d}$ determines the overall size of the dilaton potential and we demand $\lambda_d \lesssim 1$ to allow a  perturbative description. The remaining numerical prefactor is determined according to the 5D dual models~\cite{Agashe:2020lfz}.
 In \cite{Agashe:2019lhy,Agashe:2020lfz}, by considering generalized holographic models, it was shown that the parameter $\epsilon_d$ appearing in the dilaton potential (eq.~\eqref{eq:dilatonpotential}) does not need to be the same as the one that controls the large hierarchy in eq.~\eqref{eq:hierarchyepsilon}.
 In our current discussion, this additional flexibility provides us with extra freedom to achieve mild (vs severe) dilution after PT, hence making it easier to be compatible with high scale leptogenesis. Therefore, in this paper, we will treat $\epsilon_d$ as a free parameter but demand $\epsilon_d\lesssim 1$ to stay in the perturbative regime. Note that the predictions for the models where the same $\epsilon_d$ appears in eqs.~\eqref{eq:dilatonpotential} and ~\eqref{eq:hierarchyepsilon} can be simply read from our results for $\epsilon_d\sim 1/20.$

 Equating the free energy densities of the two phases, we obtain an expression for the critical temperature:
\bea\label{eq:T_c_VEV}
&F_{\rm deconfined}(T_c)=F_{\rm confined}(T_c)\approx V(\langle\phi\rangle)\nonumber\\
&\Rightarrow\frac{T_c}{\langle\phi\rangle}\approx\left(\frac{-6\epsilon_d \lambda_d}{\pi^4\, (4+\epsilon_d)}\right)^{1/4}.
\eea
We can clearly see from the above expression that $T_c$ is parametrically smaller than $\langle\phi\rangle$ for small $\epsilon_d$ and $\lambda_d$.  

{\bf The nucleation temperature} $T_n$ can be estimated as follows. A first-order PT can proceed  by nucleation of bubbles, and the bubble nucleation rate, i.e. the probability of nucleation of bubbles per unit time per unit volume,  can be estimated semi-classically as $\Gamma(T)\sim T^4 e^{-S_b}$, where $S_b$ is the action of the Euclidean bounce solution. The PT completes when this rate  in one Hubble-scale volume is comparable to the Hubble rate ($H$). Therefore, the nucleation temperature can be estimated using 
\bea
\Gamma(T_n)\sim H(T_n)^4.
\eea
It is generally hard to obtain an analytical expression for $T_n$. In this paper, we use the numerical result obtained from a bounce configuration in the dual 5D picture presented in \cite{Agashe:2020lfz}. 

{\bf The reheating temperature} $T_r$ after the completion of the PT can be estimated with the following considerations. Over the course of PT, the vacuum energy of the false vacuum converts into the energy of the thermal bath of the SM particles and the low-lying composite states of the confined phase (the low-lying KK states in the 5D picture). 
This ``reheats'' the thermal bath to a higher temperature $T_r$. Assuming this reheating happens promptly when the PT completes around $T_n$, meaning the energy density is conserved right before and after the reheating:
\bea\label{eq:T_r}
\rho_{\rm vac}+\rho_{\rm rad, before}&=&\rho_{\rm rad, after}\nonumber\\
\Rightarrow \frac{\pi^2}{8} N_c^2 T_c^4+\frac{\pi^2}{30}g_{*,\rm before} T_n^4&=&\frac{\pi^2}{30}g_{*,\rm after} T_r^4,
\eea
where we have used the equality of the free energies at $T_c$ to rewrite the vacuum energy in terms of $T_c$ (see eq.~\eqref{eq:T_c_VEV}). 
In the above equation,  $g_{*\textrm{before/after}}$ is the number of relativistic degrees of freedom in the thermal bath before/after the PT. Here 
\begin{eqnarray}\label{eq:g_star_general}
    g_{*\rm,before}=g_{*}^{\rm SM}+g_{*\rm}^{\rm CFT} ~~,~~g_{*\rm}^{\rm CFT}=\frac{45}{4}N_c^2,
\end{eqnarray}
 where $g_{*\rm}^{\rm CFT}$  denotes the degrees of freedom in the deconfined strongly coupled sector and the coefficient in front of $N_c^2$ is determined given the free energy density $F=\rho-sT$ in eq.~\eqref{eq:F_T_deconf} and the definition of $g_*$: $\rho\equiv \frac{\pi^2}{30}g_{*} T^4$.
Moreover, we assume that the SM particles dominate the energy density of the thermal bath in the confined phase at $T_r$ and thus take $g_{*,\rm after}\approx g_{*}^{\rm SM}$. This is the case as long as $T_r< \langle \phi \rangle$ because in this case contributions from typical composite resonances are Boltzmann suppressed. We have checked that this condition is satisfied in the parameter range that can avoid too much dilution for baryogenesis and have theoretical control.

Including the entropy produced during this reheating process and assuming no additional production of asymmetry during the PT, the yield for asymmetry before and after the PT are related as 
 \beq\label{eq:Y_after}
    Y_{L,{\rm after}}= Y_{L,{\rm before}}\times\eta_{\rm PT}
    \eeq
    where $\eta_{\rm PT}$ is the dilution factor due to the entropy production during the PT and is approximately given by
    \beq
    \eta_{\rm PT} \equiv\left( \frac{g_{*S,\rm before}\,T_n^3}{g_{*S\rm,after}\, T_r^3} \right). 
    \eeq
In the above equation, $g_{*S\rm,before/after}$ denotes the effective number of relativistic degrees of freedom in the thermal bath that contributes to entropy density. As mentioned before, we take $g_{*S\rm,before}=g_{*S}^{\rm SM}+g_{*S\rm}^{\rm CFT}$ with $g_{*S\rm}^{\rm CFT}=\frac{45}{4}N_c^2$, and $g_{*S\rm,after}\approx g_{*S}^{\rm SM}$ because composite states in the confined phase are heavy compared to $T_r$ and thus their contributions are negligible in the thermal bath.

\begin{figure}
    \centering
\includegraphics[width=10cm]{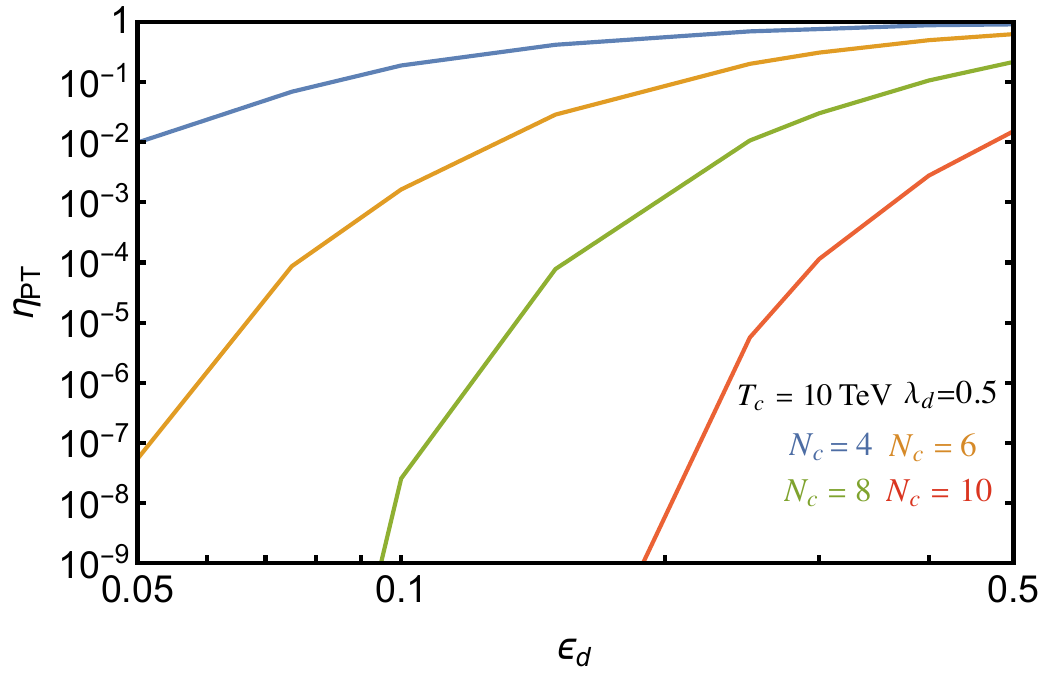}
    \caption{The dilution factor after confinement phase transition $\eta_{\rm PT}$ (eq.~\eqref{eq:Y_after}) as a function of $\epsilon_d$ for fixed $\lambda_d=0.5$ and several choices of $N_c$. Here we assume $T_c= 10$ TeV.} 
    \label{fig:Entropy_dilution}
\end{figure}

Based on the results in ref.~\cite{Agashe:2020lfz}, we plot in figure~\ref{fig:Entropy_dilution} the dilution factor  $\eta_{\rm PT}$ as function of $\epsilon_d$ for several choices of $N_c$ and fixed $\lambda_d=0.5$. 
The perturbative validity of the 5D results requires large $N_c$ and small $\lambda_d,\epsilon_d$. We see from figure~\ref{fig:Entropy_dilution} that the dilution factor can be as large as $\eta_{\rm PT}\sim 0.1(0.01)$ for large $N_c\sim 8 (10)$
and moderate $\lambda_d, \epsilon_d \sim 0.5$.

\subsection{Washout by Composite Resonances}
\label{sec:low scale washout}

After the phase transition, the strong sector confines forming composite resonances of masses set by $m_\rho$. In this phase, there can be potentially important washout effects via the \emph{on-shell} lepton-number-breaking inverse decays into composite  resonances.\footnote{We note that this is similar to the ``IR washout'' effects that was discussed in our earlier work ~\cite{Agashe:2018oyk,Agashe:2018cuf} in the context of a weakly coupled hybrid seesaw model, which at low energies reduced to an inverse seesaw model.} 
To study this washout effect, we adapt ``two-site'' approach where we only keep the lightest composite resonance (with mass of the order of the confinement scale $m_\rho$) as well as SM particles~\cite{Contino:2006nn}. 
This simplified approach is particularly suitable for us since the reheating temperature $T_r$ after the PT tends to be somewhat below the confinement scale as discussed in the previous section, and thus the on-shell production of composite resonances is Boltzmann suppressed. 
Therefore such washout effects are dominated by the inverse decays into the lightest resonance.\footnote{For the detailed relation of the couplings to neutrino mass data, there can be order one corrections due to the presence of heavier resonances. So the results of this section for the bounds on couplings and masses should be taken with potential ${\cal O}(1)$ uncertainties.}

Following a two-site approach, the Lagrangian of the neutrino sector after confinement can be written as (see \cite{Agashe:2015izu} for more details) 
\begin{eqnarray}
-\mathcal L\supset y \Psi^c HL+m_{\Psi} \Psi^c \Psi+\mu_\Psi \Psi\Psi,
\end{eqnarray}
where $H$ is the SM Higgs, $L$ is the SM lepton doublet, $y$ is the Yukawa coupling, $\Psi$ is the field corresponding to the lightest of resonances created by acting with operator $O$ on the vacuum,
and $\Psi^c$ is the Dirac partner of $\Psi$. 
$m_{\Psi}$ is the Dirac mass of the resonance and $\mu_\Psi$ is the (small) lepton number violating mass, i.e.~Majorana mass, which originates from the operator $O^2$ in eq.~\eqref{eq:CFTatmrho}. We can estimate the size of $m_{\Psi}$ and $\mu_\Psi$ as 
\begin{eqnarray}
m_\Psi\sim m_\rho~,~~\mu_\Psi\sim\frac{\bar{\lambda}^2(\Lambda) m_\Psi^2}{{g_\rho^2} m_N}\left(\frac{ m_\Psi }{m_N}\right)^{2\Delta-5} \ll m_\Psi.
\end{eqnarray}
Therefore, the neutrino sector has an effective  inverse-seesaw-like description in the confined phase \cite{Agashe:2015izu}.

Similarly to the case of leptogenesis in TeV scale inverse seesaw models, the generation of asymmetry from decay and inverse decay of $\Psi$ or $\Psi^c$ are generically negligible.\footnote{We note that degeneracy among different generations of $\Psi$'s could resonantly enhance the asymmetry~\cite{Agashe:2018cuf,Dichtl:2023xqd}. However, in this work, we assume generic flavor structure, hence we do not consider resonance enhancement effects.}  The dominant effect of $\Psi$ and $\Psi^c$ is then to wash out pre-existing asymmetry and the most relevant process is the scattering $H L \to \bar{H} \bar{L}$ mediated by on-shell $\Psi$ and $\Psi^c$. 
Defining a time variable $z\equiv m_\Psi/T$, one needs to solve the relevant Boltzmann equations starting from $z_i=m_\Psi/T_r$. Following the results in ref.~\cite{Agashe:2018cuf}, the asymmetry as a function of $z$ can be written as
\begin{eqnarray}\label{eq:Y_z_general}
Y_{L}(z)=Y_{L}(z_i) \; \mathrm{exp}\left[-\frac{66}{79\pi^2}K^{\rm eff}_\Psi f(z_i,z)\right],
\end{eqnarray}
where $Y_{L}(z_i)$ with $z_i\equiv m_{\Psi}/T_r$ is the primordial asymmetry right after the phase transition, which is $Y_{L,\rm after}$ in eq.~(\ref{eq:Y_after}).  $K_{\Psi}^{\rm eff}$ is the effective washout parameter defined as \cite{Blanchet:2009kk} 
\begin{eqnarray}
K_{\Psi}^{\rm eff}= \left. \left(\frac{\mu_\Psi}{\Gamma_{\Psi}}\right)^2 \left( \frac{\Gamma_\Psi}{H} \right) \right\vert_{T=m_\Psi},
\end{eqnarray}
for $\mu_\Psi\ll\Gamma_\Psi$. The appearance of a factor $\propto \mu_\Psi^2$ in the above equation can be understood by noting that in the limit $\mu_\Psi \rightarrow 0$, there is no lepton number violation and hence no washout.\footnote{ The absence of a linear term in $\mu_\Psi$ can be understood by noting that the rates are obtained from the square of the amplitudes and there is no contribution to the amplitude at the zeroth order in $\mu_\Psi$ since the washout processes involve initial and final states with different lepton numbers.  
The two powers of $\Gamma_\Psi$ in the denominator comes from the resonant enhancement for on-shell s-channel scattering involving the nearly degenerate pseudo-Dirac pair $\Psi$: the mass splitting ($\mu_\Psi$) between the pseudo-Dirac pair is smaller than their width $\Gamma_\Psi$. } The temperature dependence is encoded in $f(z_i,z)$:
\bea\label{eq:washout_f}
f(z_i,z)=\int_{z_i}^z dz' z^{\prime 3} \mathcal K_1(z'),
\eea
where $\mathcal{K}_1$ is the modified Bessel function of the first kind.

Since the baryon number freezes out when the EW sphaleron process gets out of the equilibrium at a temperature $T_{\rm sph}$, the effect of the washout can be estimated by evaluating eq.~\eqref{eq:washout_f} at $z=z_{\rm sph}\equiv m_\Psi/T_{\rm sph}$.
Depending on whether the value of $T_r$, which we will  estimate below in eq.~\eqref{eq:estimateTr}, i is below or above $T_{\rm sph}$ two different scenarios are possible.  First possibility is that the universe is reheated to  $T_r<T_{\rm sph}$, the baryon asymmetry $Y_B$ is simply frozen at $T_r$, even if the lepton asymmetry can still be washed out.
 The second possibility, which as we will see below occurs more generically in the allowed parameter space, is that the universe is reheated to $T_r>T_{\rm sph}$. 
For $z_{\rm sph} > z_i \gtrsim 1$ we get 
\beq
f(z_i,z_{\rm sph}) \approx \sqrt{\frac{\pi}{2}} \left( z_i^{5/2} e^{-z_i} - z_{\rm sph}^{5/2} e^{-z_{\rm sph}} \right)
\eeq
which manifests the Boltzmann suppression for $T_r< m_\Psi$ as we anticipated (similar results have also been obtained in ref.~\cite{Dichtl:2023xqd}). Therefore, it is convenient to define the IR washout suppression factor $\eta_{\rm IR}$ as
\begin{eqnarray}\label{eq:eta_IR}
   \eta_{\rm IR}\equiv\mathrm{exp}\left[-\frac{66}{79\pi^2}K^{\rm eff}_\Psi f(z_i,z_{\rm sph})\right],
\end{eqnarray}
such that we can write the asymmetry after IR washout at $z=z_{\rm sph}$ as
\begin{eqnarray}
  Y_{L}(z_{\rm sph})=Y_{L}(z_i) \; \eta_{\rm IR}.
\end{eqnarray}

Given the  double-exponential sensitivity of the washout to $z_i$, it is important to provide a reasonable estimate for this quantity. 
For this we use the concrete 5D model described in the previous section. 
It is useful to recall that in the absence of large brane-localized kinetic terms, mass of the first fermion KK excitation is $m_\Psi \sim \pi \langle \phi \rangle $ \cite{Gherghetta:2000qt}. Using this and eqs.~\eqref{eq:T_c_VEV} and \eqref{eq:T_r} we obtain for $N_c \gtrsim 3$ 
\beq \label{eq:estimateTr}
z_i^{-1}=\frac{T_r}{m_{\Psi}}\sim 0.05 (\epsilon_d \lambda_d)^{1/4} \sqrt{N_c}.
\eeq
For $\epsilon_d, \lambda_d\lesssim 0.5$ and $N_c\sim 10$, $z_i=m_\Psi/T_r$ is generically $\gtrsim 10$, and $z_i$ can be as large as $z_i\sim 30$ for $\lambda_d \sim 0.5, \epsilon_d \sim 0.05$ and $N_c\sim 3$.
Considering the values of $m_\Psi$ above a few TeV as allowed by phenomenological constraints, we see that we are generically in the regime $z_i>z_{\rm sph}$. 
The estimate $z_i\gtrsim 10$ implies a large suppression of the washout, primarily due to $f(z_i, z_{\rm sph})$ being exponentially small to the level of $10^{-2}$ or smaller.

Finally, we discuss the IR washout factor $K_{\rm eff}$. In \cite{Agashe:2018cuf}, it was shown that the size of $K_{\rm eff}$ can be estimated to be
\beq 
K_{\Psi}^{\rm eff} \sim \frac{16 \pi}{\sqrt{g_*}} \frac{m_{\rm Pl} m_\nu^2 m_\Psi}{v^4 y^6}.
\eeq
The Yukawa coupling $y$ can be estimated in terms of the strength of the coupling among composite states $g_\rho$ and the elementary-composite mixing angle for $L$ to be 
$y \sim g_\rho s_{\nu}$.
One can obtain the anarchic neutrino mass and mixings by having anarchic $s_{\nu}$.\footnote{In models where $s_L=s_{\nu}$, the charged lepton mass hierarchy is obtained by having hierarchies in the compositeness of the right-handed leptons, $s_e$.
In these models where the same elementary-composite mixing enters into charged lepton and neutrino masses, we can estimate the range of $y$ by considering the tau Yukawa coupling,  
\beq
\frac{y}{y_\tau}\sim \frac{g_\rho s_{L}}{g_\rho s_{L}^\tau s_e^\tau}\sim \frac{1}{s_e^\tau} >1.
\eeq
Therefore in these models we have $y_\tau<y<g_\rho$.   }

\begin{figure}[h!]
    \centering
\includegraphics[width=7cm]{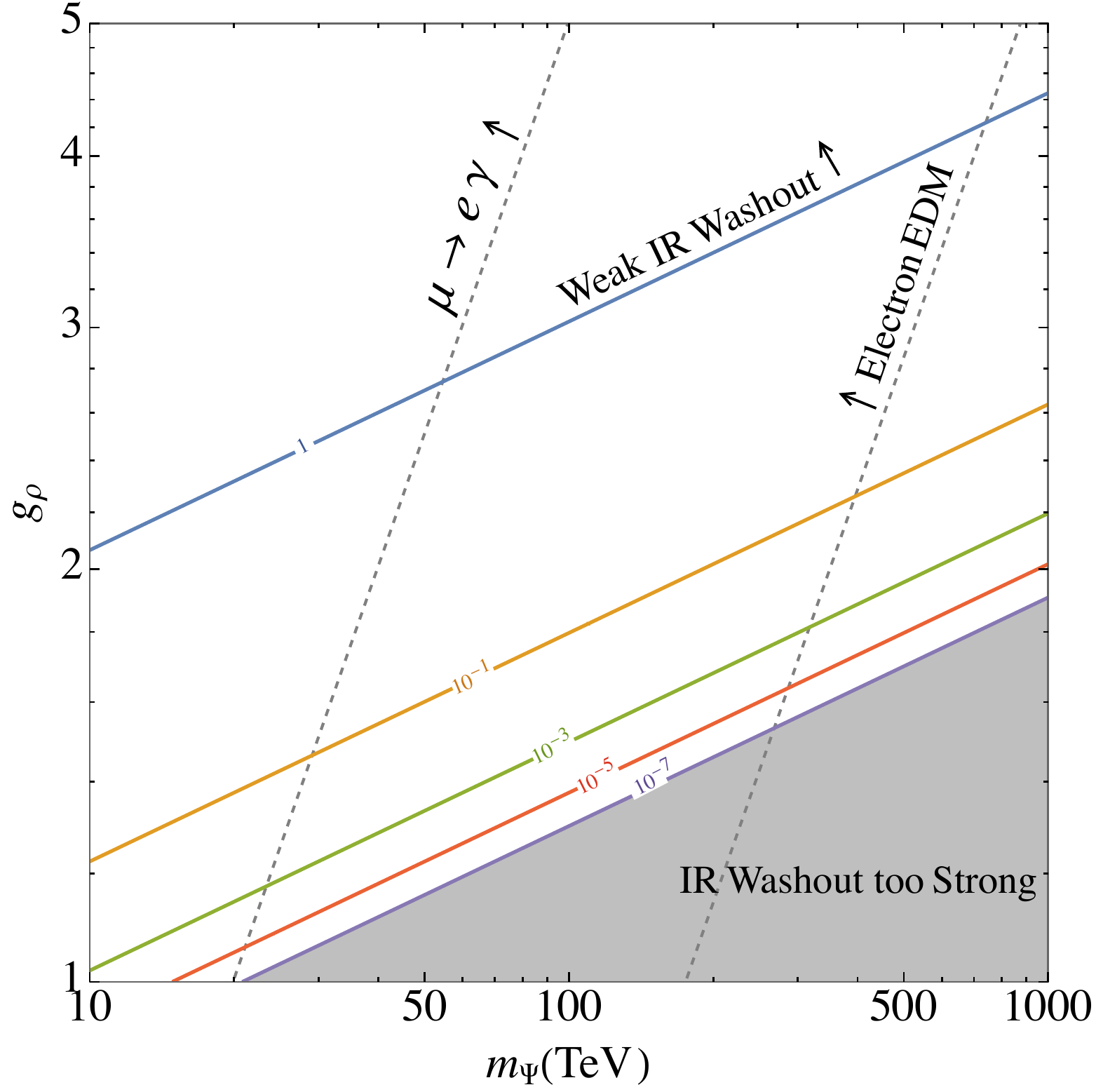}
\includegraphics[width=7.25cm]{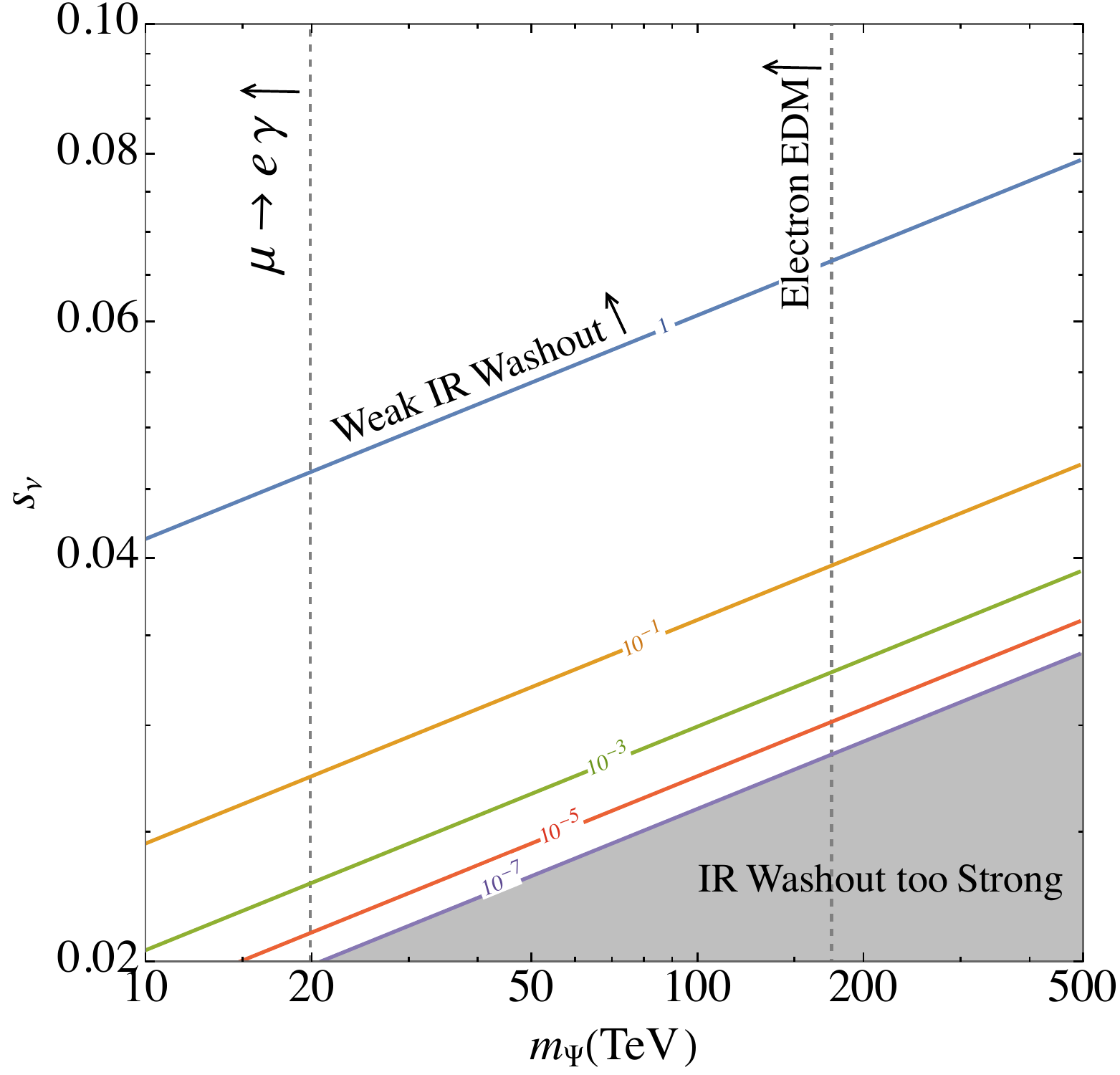}

    \caption{Contours of IR washout efficiency $\eta_{\rm IR}$ in eq.~\eqref{eq:eta_IR} (colored lines) with $m_\nu=0.05$ eV and $z_i=10$ in eq.~(\ref{eq:Y_z_general}). In the left panel, we present it in the $(m_{\Psi},g_\rho)$ plane with $s_\nu=0.02$. In the right panel, we show it in the $(m_{\Psi},s_\nu)$ plane fixing $g_\rho=1$.  The gray shaded region is not allowed because the IR washout from composite resonances is too strong. The IR washout is weak above the blue line and thus the primordial asymmetry is not affected. Assuming generic partial compositeness structure, the region above gray dashed lines is ruled out by $\mu\to e\gamma$ or electron EDM measurements. These bounds can however be relaxed by imposing flavor symmetries on the structure. }
    \label{fig:IR_washout}
\end{figure}

In Fig.~\ref{fig:IR_washout} we show the IR washout suppression $\eta_{\rm IR}$, which reflects 
the suppression of asymmetry by the on-shell inverse decay processes. In the gray shaded region the IR washout is so significant even an ${\cal O}(1)$ initial asymmetry is not sufficient to produce the observed value. The region defines a lower bound on $s_\nu$:
\beq\label{IRwashoutcond}
s_{\nu}\gtrsim 0.02 \left( \frac{1}{g_\rho} \right) \left( \frac{f(z_i, z_{\rm sph})}{0.01} \right)^{1/6} \left( \frac{m_\Psi}{10 {\rm TeV}} \right)^{1/6},
\eeq
which can be compared with the condition of eq.~\eqref{eq:bound_offshell_wo_1} from $\Delta L=2$ washout before the PT. These conditions both prefer a sizable compositeness for neutrinos and lead to parametrically different but numerically comparable constraints.

\section{Finally: BAU, Neutrino Mass, and Phenomenology} 
\label{sec:baryon_asymmetry}
In this section, we combine our results to obtain the observed BAU and discuss direct as well as indirect constraints on the model. Taking into account all these considerations we then provide representative benchmark models.

\subsection{The observed baryon asymmetry of the universe}

Below the temperature where the washout by the $\Psi$ resonances and the EW sphaleron processes get out of equilibrium, the $B-L$ charge is distributed into $B$ and $L$, which separately become conserved quantum numbers.\footnote{We remind the reader that as we discussed in the beginning of section~\ref{sec:HighscaleBE},  $B-L$ is the relevant slowly-varying charge as $B+L$ is violated by sphaleron process which are in equilibrium over a wide range of temperatures above the weak scale. However for ease of notation and language we simply referred to the relevant slowly varying charge as lepton number. In this section we carefully use the appropriate quantum numbers involved.}
Assuming that when the EW sphalerons freeze out (and after the confining PT) the EW symmetry is already broken such that the number of relativistic degrees of freedom is that of the SM minus the contribution from top quarks, we have~\cite{Harvey:1990qw,Inui:1993wv} 
\begin{equation}
    Y_B = \frac{30}{97} Y_{B-L}. 
    \label{eq:B_B-L_relation}
\end{equation}

The final formula for the baryon asymmetry generated in our models is given by  
\bea\label{eq:Y_B_general}
Y_{B}=\frac{30}{97}\times  Y_{B-L}^{\rm UV}\times \eta_{\rm PT}\times \eta_{\rm IR}
\eea
where $Y_{B-L}^{\rm UV}$ is the asymmetry generated at high scale from $N$ decay, which we take to be
\bea
Y_{B-L}^{\rm UV}=
\begin{cases}
    \epsilon \sqrt{K} & \rm{(Weak~Washout)}\\
    c \, \epsilon/(g_* K) & \rm{(Strong~Washout)}
\end{cases}
\eea
In the strong washout regime, the generated asymmetry is independent of initial conditions and is taken from eq.~\eqref{eq:strong_wo_UV}, assuming the overall factor of order unity is one. While in the weak washout regime, as we discussed in section \ref{sec:weakwashout}, the result in general depends on the assumed scenario and initial conditions. Here we use the expression of eq.~\eqref{eq:Y_L estimate weak wo} corresponding to a scenario where $N$ dominated the energy density and led to a result less sensitive to initial conditions, again with numerical coefficient taken to be one. The CP asymmetry parameter $\epsilon$ from $N$ decay is in eq.~\eqref{eq:epsilon_simplified}, while the washout factor $K$ is defined in eq.~\eqref{eq:defineK}. $\eta_{\rm PT}$ is the entropy dilution factor due to the reheating after the confining phase transition (see eq.~(\ref{eq:Y_after})). $\eta_{\rm IR}$ denotes the suppression due to washout in the IR from the inverse decay to the composite resonances after the phase transition (see eq.~(\ref{eq:Y_z_general})).

\begin{table}
\hspace*{-0.2cm}
\begin{tabular}{|c|c|c|c||c|c|c||c|c||c|c|c|} 
 \hline 
  &\multicolumn{3}{c||}{High scale genesis} & \multicolumn{3}{c||}{Confinement PT}  &\multicolumn{2}{c||}{Low scale} &\multicolumn{3}{c|}{Derived quantities} \\
 \cline{2-12}
  &$m_N$(GeV) & $\Delta$ & $s_\nu$ & $\epsilon_d$ & $\lambda_d$ & $N_c$ & $g_\rho$ & $m_\Psi$ (TeV) & $Y_{B-L}^{\rm UV}$ & $\eta_{\rm PT}$ & $\eta_{\rm IR}$\\ \hline \hline

BP1 &$7.5\times 10^{12}$ &2.35 &$0.1$ &0.36 &0.5 & 10 &3.0 &10 &$2.9\times10^{-7}$ &$10^{-3}$ &$1$\\
\hline
BP2 &$10^{15}$ &2.20 &0.02 &0.30 &0.5 & 10 &1.2 &10 &$2.9\times10^{-5}$ &$10^{-4}$ &$10^{-1}$\\
\hline
BP3 &$10^{12}$ &2.13 &0.1 &0.38 &0.5 &8 &4.0 &3 &$2.9\times10^{-9}$ &$10^{-1}$ & 1\\
 \hline
 \end{tabular} \\
 \caption{Benchmark points for conformal leptogenesis that can achieve the observed BAU. The general formula for $Y_{B}$ is given in eq.~\eqref{eq:Y_B_general}. For BP1, the UV asymmetry ($Y_{B-L}^{\rm UV}$) is generated in the strong washout regime, while the other two are in the weak washout regime. For all cases, we set the UV cut-off as $\Lambda=10^{16}$ GeV, the confinement scale $m_\rho=m_\Psi$ and SM neutrino mass $m_\nu=0.05$ eV.}
 \label{tab:benchmark}
\end{table}

In table~\ref{tab:benchmark}, we provide several benchmark points that can achieve the observed BAU $Y_{B}^{\rm obs}\approx 9\times 10^{-11}$~\cite{Planck:2018vyg}, as well as the SM neutrino mass $m_\nu=0.05$ eV. As can be seen from table~\ref{tab:benchmark}, conformal leptogenesis allows for a broad range of $m_N$ and lepton asymmetry generated from UV. 
As we can see in figure~\ref{fig:Highscale_asymmetry}, in the strong washout regime, which is the more predictable regime given the asymmetry being independent of initial conditions, the maximum UV asymmetry that can be achieved is $Y_{B-L}^{\rm UV}\sim 10^{-6}$. Therefore in this regime, the maximum acceptable suppression due to the PT and the subsequent inverse decay is of order $\mathcal{O}(10^{-4})$. The BP1 in table \ref{tab:benchmark} shows a representative point in this regime. 
We also note that there is a maximum $m_N$, beyond which leptogenesis is not successful in the strong washout regime. This upper bound on $m_N$ is given roughly by $m_N\lesssim 10^{14}$ GeV as can be seen from figure~\ref{fig:Highscale_asymmetry}.

For larger $m_N$, successful leptogenesis would necessarily happen in the weak washout regime. Furthermore, the weak washout regime also allows for somewhat larger UV asymmetry than maximal allowed value in the strong washout regime. BP2 of table~\ref{tab:benchmark} demonstrates this point by showing a representative set of parameters realizing the weak washout regime with $m_N\sim 10^{15}$ GeV.

For BP1 and BP2 we have chosen  $m_\Psi =10$ TeV, which are clearly safe from the current bounds from direct searches (see next section). Instead, the last benchmark BP3 shows a point with $m_\Psi=3 $ TeV which can be potentially probed at the LHC (see, e.g., \cite{Agashe:2016ttz,Agashe:2017ann}). Furthermore BP3 leads to a deviation in the coupling of neutrinos to $Z$ of $\delta g_\nu / g_\nu \sim 10^{-3}$, which is not far from the current bound (see again next section).

\subsection{Other phenomenological constraints}

The lower the compositeness scale $m_\rho$, the more natural is the CH model from the viewpoint of the hierarchy problem. Current bounds from direct searches of resonances require $m_\rho$ to be above a few TeV. Stronger constraints arise from flavor and CP violation. In this subsection we collect a discussion of the constraints that are directly related to our mechanism of leptogenesis.

\subsubsection{Probing the neutrino compositeness} 

An interesting requirement imposed by the $\Delta L=2$ washout effects (eq.~\eqref{eq:bound_offshell_wo_1}) as well as the IR washout (see \eqref{IRwashoutcond}) is the need for a rather large degree of compositeness of neutrinos, i.e. a sizable $s_\nu$. A large $s_\nu$ can be probed from precision measurements  of the couplings of the LH leptons to the SM EW gauge bosons. The deviation can be estimated as
\beq
\frac{\delta g_\nu}{g_\nu} \sim s_\nu^2 \left( \frac{g_\rho \, v} {m_\rho} \right)^2
\eeq
As already explained below eq. \eqref{matchingOHL}, in generic CH scenarios a similar constraint would apply to both the charged LH lepton and the neutrino. However using a custodial protection for coupling to $Z$ ---as was originally employed  in ref. \cite{Agashe:2006at} to protect  coupling of $b$ quarks to $Z$--- it is possible to relax the deviations for either the charged lepton or the neutrino coupling to $Z$, but not both simultaneously. Given that the constraints on the charged leptons are stronger, we assume a custodial protection for the couplings of charged leptons and use the above estimate for the deviation of the neutrino couplings. This can be probed from the precision measurements of the {\em in}visible decay width of the $Z$ boson, which is currently measured with a precision of $\mathcal{O}(0.3 \%)$ \cite{ParticleDataGroup:2024cfk}, resulting in a bound $\delta g_\nu / g_\nu \lesssim 0.15 \%$.
In the future, new colliders such as the FCC-ee are expected to improve the precision of the $Z$ invisible decay width measurement by a factor of 100 compared to the current bound (see, e.g. \cite{Blondel:2021ema})

\subsubsection{$\mu\to e$ transitions and electric dipole moments}

As already anticipated above, important constraints on CH models come from flavor CP observables. The processes that are more directly influenced by our parameters are $\mu\to e$ transitions and the electron electric dipole moment.

In figure~\ref{fig:IR_washout}, we present the current constraints from $\mu\to e\gamma$~\cite{MEG:2016leq} and the electron EDM~\cite{Roussy:2022cmp} in the case of generic Composite Higgs models. These are shown with gray dashed lines with associated labels. The region of parameter space indicated with an arrow sign is ruled out by the respective search. Yet, it is  important to mention that these bounds may be considerably relaxed by introducing suitable flavor symmetries (see for e.g.~\cite{Frigerio:2018uwx,Glioti:2024hye}).

\section{Conclusions}
\label{sec:conclusion}

The origin of the observed baryon asymmetry is one of the outstanding open questions of particle physics. Several mechanisms for ``baryogenesis'' have been proposed in the past. Among them, the most predictive are certainly those embedded in frameworks which also address other open problems. In this paper we considered the framework of composite Higgs models with partially composite fermions, including partially composite right-handed neutrinos. Here an approximate strongly-coupled conformal strong Higgs dynamics, often simply called CFT, breaks the SM flavor symmetry via linear couplings between the fundamental fermions and CFT operators, and naturally generates a composite Higgs boson with a mass parametrically smaller than the UV cutoff. Those models are popular solutions of the Planck-weak hierarchy and in principle can also provide an explanation of the flavor hierarchy among the SM fermion masses.

In our setup the strong Higgs dynamics, i.e. the CFT, is assumed to respect the lepton-number. However, its coupling to a fundamental heavy Majorana singlet explicitly breaks it. Upon integrating-out that, an irrelevant small amount of lepton-number breaking is injected into the theory. After the approximate CFT develops a mass gap, the explicit breaking manifests itself as tiny Majorana mass splittings for the resonances (which would otherwise be purely Dirac). The exchange of pseudo-Dirac resonances finally generates small SM neutrino masses, in a way that is reminiscent of the inverse seesaw mechanism. We thus see that what one may have expected to be a standard seesaw setup for neutrino masses, in reality induces a mechanism that is a sort of hybrid between high-scale and inverse seesaw. How does leptogenesis work in such a setup? This is the question we addressed in the present paper.

Analogously to neutrino masses, we found that also leptogenesis does not go through as in the standard scenario. Two important qualitative differences appear. Firstly, being a finite temperature phenomenon, leptogenesis in our setup takes place in an era in which the singlet decays into the plasma of a deconfined strong dynamics, as opposed to decaying into ordinary weakly-coupled leptons and Higgses as in the standard scenario. Therefore, care should be paid when describing such ``conformal leptogenesis''. The formalism we adopted is very general and could be of interest in other theoretical applications. Secondly, as the temperature decreases the CFT plasma turns into a theory of interacting particles, and the phase transition as well as the resonances that result from it do have an impact on the present-day asymmetry. None of these effects are present in standard leptogenesis.

Our main conclusion is that there exist vast regions of parameter space in which the observed baryon asymmetry is reliably calculable and of the correct order of magnitude. In addition, characteristic testable signatures are expected. Perhaps the most novel phenomenological implication is a lower bound on the strength with which neutrinos couple to the Higgs dynamics. In generic composite Higgs scenarios with partial compositeness, a strong mixing between neutrinos and the CFT would imply a similar coupling for the charged leptons, which would be severely disfavored by data. Yet, we argued that reasonable generalizations can split the couplings of neutrinos from those of their charged electroweak partners and that the resulting scenarios can be consistent with all known data. Future precision tests, for example $Z$-pole observables at FCC-ee, will be capable of exploring these scenarios further. Other signatures that characterize our models include the production of singlet fermionic resonances with sizable Yukawa couplings and masses at the TeV scale,  potentially testable at the LHC and future colliders (see e.g. \cite{Agashe:2016ttz,Agashe:2017ann}). Finally, the strongly first order confining phase transition can lead to the production of gravitational waves. Our results suggest that the transition be not over-supercooled, an observation that correlates with a faster completion of the bubble percolation, directly constraining the allowed spectrum of the gravitational wave signal.

\section*{Acknowledgements}

We thank Riccardo Rattazzi and Emilio Trevisani for discussions,
and Matthew Walters for pointing to us the relevant literature on CFT at finite temperature.
The work of KA is supported by NSF Grant No.~PHY-2210361 and by the Maryland Center for Fundamental Physics. PD is supported by DOE grant DOE-SC0010008. The work  of ME is  supported by the Swiss National Science Foundation under contract 200020-213104. ME also acknowledges the hospitality of the CERN theory group where part of this work was completed.
CSF acknowledges the support by Fundacão de Amparo à Pesquisa do Estado de São Paulo (FAPESP) Contract No. 2019/11197-6 and Conselho Nacional de Desenvolvimento Científico e Tecnológico (CNPq) under Contract No. 304917/2023-0.
The work of SH is supported by the National Research Foundation of Korea (NRF) Grant RS-2023-00211732, by the Samsung Science and Technology Foundation under Project Number SSTF-BA2302-05, and by the POSCO Science Fellowship of POSCO TJ Park Foundation. The work of LV was partly supported by the Italian MIUR under contract 202289JEW4 (Flavors: dark and intense), the Iniziativa Specifica “Physics at the Energy, Intensity, and Astroparticle Frontiers” (APINE) of Istituto Nazionale di Fisica Nucleare (INFN), and the European Union’s Horizon 2020 research and innovation programme under the Marie Sklodowska-Curie grant agreement No 860881-HIDDeN.

\appendix

\section{ Boltzmann Equations near Equilibrium } \label{app:nearequilibrium}

In this appendix we briefly review the formalism that was developed in refs.~\cite{Bodeker:2014hqa, Bodeker:2017deo} for the standard type-I seesaw leptogenesis and then apply it to the models we are considering to obtain the Boltzmann equations as an expansion near equilibrium. This also allows us to express the rates appearing in the evolution equations in terms of  correlation functions of the CFT.

\subsection{General formalism}

We start by reviewing the general formalism, following ref.~\cite{Bodeker:2014hqa}. 
The basic idea is that after identifying the right set of variables that vary only on long distances and times (compared to the time/length scales of the microscopic theory as well as those set by the temperature), one can write an effective description for these ``slow variables'' only. In a homogeneous system, the slow variables are simply the (densities of) approximately conserved charges. 
For a set of slowly varying ``charges'' $Q_a$ and for small deviations from equilibrium,  we can write the linearized equations as 
\beq \label{eq:generallinearizedBE}
\mathcal{D}_t Q_a=-\Gamma_{ab} \left( Q_b -Q_b^{\rm eq} \right).
\eeq
Here $Q_b^{\rm eq}$ denotes the equilibrium value of the charges, 
$\mathcal{D}_t$ is a generalized time derivation operator, which for the case at hand is given by $\mathcal{D}_t=\frac{d}{dt}+3 H$, and $\Gamma_{ab}$ are rates to be identified below.

In the present problem the slowly-varying variables are $n_N$, the number density of $N$, and $n_L$, the lepton number density. By restricting to the regime of non-relativistic $N$ and  $T\ll m_N$, the linearized equation can be written near equilibrium in the following form:
\bea
\mathcal{D}_t n_N && = - \Gamma_{\rm D}(n_N- n_N^{\rm eq})%
\\
\mathcal{D}_t n_L && = -\Gamma_W n_L+ \Gamma_S (n_N- n_N^{\rm eq})
\eea
where the RHS of these equations are expansions in powers  $n_a - n_a^{\rm eq}, \; a= N {\rm \; or \;}  L$, with $ n_L^{\rm eq}=0$. 
We have neglected terms that are higher order in $n_i-n_i^{\text{eq}}$ in both equations. In particular, a possible term linear in $n_L$ in the first equation is dropped  working in the regime where $n_L$  is formally smaller than $n_N$ as $n_L$ is suppressed by the CP violating spurions while $n_N$ is not. 

The coefficients denoted by $\Gamma_D$, $\Gamma_W$ and $\Gamma_S$ are in general temperature-dependent and can be computed in the microscopic theory. In Appendix \ref{app:Washout_from_Kubo} we derive the washout rate $\Gamma_W$ and discuss its relation with the $N$-decay rate $\Gamma_D$, whereas $\Gamma_S$, which corresponds to the rate for the (asymmetry) source term, will be derived in Appendix \ref{app:source}. 
As we will see, all relevant rates can be expressed in terms of certain correlation functions of the CFT. Their zero temperature limit is captured by our computation in section~\ref{subsec:computeepsilonGamma}. Finite temperature corrections are argued to be suppressed by powers of $T/m_N$ in Appendix \ref{app:finiteT}.

\subsection{Washout coefficient in terms of the CFT correlators} \label{app:Washout_from_Kubo}

We now present the derivation of the washout rate $\Gamma_W$, using the fluctuation-dissipation theorem. Following ref.~\cite{Bodeker:2014hqa}, it is convenient to express  $\Gamma_W$ in terms of correlators of $\dot{Q_L}$, the time derivative of the charge $Q_L$ via
\beq \label{eq:washoutrateQdot}
\Gamma_{W}=  \frac{\xi^{-1}}{2 V T^2 } \lim_{\omega \to 0} \frac{1}{\omega} \int dt e^{i \omega t} \left\langle \left[ \dot{Q_L}(t),\dot{Q_L}(0) \right] \right\rangle_T
\eeq
where the latter correlator is evaluated at the zeroth order in $\lambda$ and $\langle\cdots\rangle_T$ indicates the thermal average. The order one dimensionless coefficient $\xi$ is the susceptibility normalized by $T^2$, and is defined as
\beq
\xi= \frac{1}{V T^3} \langle Q_L^2 \rangle_T .
\eeq
$\xi$ can be determined by identifying the fast processes that are in thermal equilibrium for a given temperature $T$.

For the case at hand, an explicit expression for $\dot{Q_L}$ can be obtained from Noether's current, which gives
\beq
\dot{Q_L}=i \int d^3x \left[ \lambda_{i \alpha} \overline{\Psi}_{N_i} O_\alpha - \lambda^*_{i \alpha} \overline{O}_\alpha \Psi_{N_i}  \right].
\eeq
Eq.~\eqref{eq:washoutrateQdot} can be explicitly evaluated and the result is 
\beq \label{eq:washoutspectral1}
\Gamma_{W}=- \frac{\xi^{-1}}{  T^2} \lambda_{i \alpha} \lambda^*_{i \alpha} \int \frac{d^3k}{(2 \pi)^3} \frac{f_F^\prime (E_i)}{2 E_i} \rm{Tr}\left[ \slashed{k} (\tilde{\rho}_{\alpha }(k) +\tilde{\rho}_{\alpha }(-k) ) \right]\big|_{k^0=E_i}
\eeq
where $f_F$ is the Fermi-Dirac distribution, the prime denotes the derivative with respect to its argument, and $E_i=\sqrt{\bold{k}^2+m^2_{N_i}}$. The trace is over the spinor indices and $\tilde{\rho}_\alpha$ is the spectral function of the operator $O_\alpha$, 
\begin{align}
\tilde{\rho}_\alpha(k)
&=\frac12 \int d^{4}x e^{i k \cdot x} \langle \{ O_\alpha(x) , O^\dagger_\alpha (0) \} \rangle_T,
\end{align}
where the spinor indices have been omitted. The latter quantity can be shown to coincide with the discontinuity of the  thermal 2-point function (see \cite{Laine:2016hma}). Hence we find that $\Gamma_W$ is essentially controlled by the thermal average of \eqref{propagator}. In the non-relativist limit, $T \ll m_{N_1}$, we can neglect the power law corrections estimated in Appendix \ref{app:finiteT}. At leading order in the expansion in powers of ${T}/m_{N_1}$ we can thus approximate the spectral function by its zero temperature expression 
\beq \label{eq:specdenszeroT}
\tilde{\rho}_\alpha(k)={\cal C}_\alpha {\rm sign}(k_0) P_L\slashed{k} (k^2)^{\Delta_\alpha-5/2}
\eeq
The trace in the eq.~\eqref{eq:washoutspectral1}, when evaluated at the on-shell four-momenta $k$ with $k^2=m_N^2$, becomes $k$-independent and can be taken out of the integral. The remaining integral can be done independently of the model parameters. Furthermore, in the non-relativistic limit $f_F^\prime (E_i)\approx -f_F(E_i)/T$, so the integral is a familiar thermal average. The result reduces to 
\bea
\Gamma_W\approx  \xi^{-1} \frac{n_N^{\rm eq}(T)}{T^3} \langle\Gamma_N\rangle_T.
\eea
This is precisely the expression that appears in \eqref{eq:BoltzeqAsymLV} provided we replace the decay width with its thermal average and identify $\xi\approx c/6$.\footnote{This is in fact the case as in the grand-canonical ensemble, the charge density is given by the first derivative of the pressure with respect to the chemical potential, while susceptibility is given by its second derivative. We simply have $\xi=\frac{1}{T^2} (\partial n_L/\partial \mu)|_{\mu=0}= c/6$. }

Similarly, the production rate of $N$ can be written in terms of the CFT spectral function as \cite{Asaka:2006rw,Laine:2011pq} 
\beq \label{eq:productionspectral}
\gamma_{{\rm prod}, N}\equiv\Gamma_{\rm D} n_N^{\rm eq}= \lambda_{i \alpha} \lambda^*_{i \alpha} \int \frac{d^3k}{(2 \pi)^3} \frac{f_F(E_i)}{2 E_i} {\rm{Tr}} \left[ \slashed{k} (\tilde{\rho}_{\alpha }(k) +\tilde{\rho}_{\alpha }(-k) ) \right] \big|_{k^0=E_i},
\eeq
which using the zero-temperature spectral function gives $\Gamma_D=\langle \Gamma_N \rangle_T$, as in \eqref{eq:BoltzmanneqNLV}. 
More generally,  considering also the finite temperature corrections to the spectral function, by comparing this equation with eq.~\eqref{eq:washoutspectral1} and noting that  $f_F^\prime (E_i)\approx -f_F(E_i)/T$ for $T \ll m_{N_i}<E_i$,
it is readily seen that 
(at leading order in the slowly varying densities, in the non-relativistic regime, and at leading non-trivial order in $\lambda$) 
washout and production rates are proportional to each other, related by $\Gamma_W= \frac{\xi^{-1}}{T^3}\gamma_{{\rm prod},N}$. 

\subsection{The source term}
\label{app:source}

In this subsection we provide an expression of the source rate $\Gamma_S$ for the particular case in which $m_{N_1}\ll m_{N_{2,3}}$. We follow ref. \cite{Bodeker:2017deo},  where it was derived for the standard Type-I leptogenesis. For the purpose of studying the asymmetry generated by $N_1$ we can integrate out the heavier $N$'s. This gives rise to  an effective interaction 
\beq
\mathcal{L} \supset
\sum_{ i\neq1} \frac{1}{2m_{N_i}}\lambda_{i\alpha} \lambda_{i \beta} O_{\alpha}O_\beta +{\text{h.c.}}.
\eeq
The authors of \cite{Bodeker:2017deo} find that the CP violating rate $\Gamma_S$ can be written in terms of three-point correlators with the above operator inserted at one point and $O^\dagger_\alpha$ at the other two points,   $\langle [O_{\alpha}^\dagger(x_1) O_\beta^\dagger(x_2)][ O_{\gamma}(0)O_\sigma(0)] \rangle_T$, with $[\cdots]$ indicating the contraction of the spinor index. We denote the corresponding generalized (momentum-space) spectral function, as defined in \cite{Bodeker:2017deo}, by $\hat{\rho}_{\alpha \beta \delta \sigma}(k_1,k_2)$, with $k_1$ and $k_2$ being the 4-momenta conjugate to $x_1$ and $x_2$ respectively. With this notation the collision term associated to the source is given by 
\beq
\gamma_S=\int \frac{d^3 k}{(2 \pi)^3} [f_N (E_1)-f_F(E_1)]  \frac{\hat{\rho}_{\alpha \beta \gamma \sigma}(-k, k)}{4 E_1} m_{N_1} \sum_{i \neq 1}\frac{1}{m_{N_i}}{\text{Im}}[\lambda_{i\gamma} \lambda_{i \sigma} \lambda^*_{1 \alpha} \lambda^*_{1 \beta}].
\eeq
At leading order in large-$N_c$ counting, the spectral function factorises into spectral functions corresponding to the two point functions and, for small temperatures, the latter can be approximated by their zero temperature expressions. Making these approximations we obtain 
\beq
\hat\rho_{\alpha \beta \gamma \sigma}(k_1,k_2)\approx  -(\delta_{\alpha \gamma} \delta_{\beta \sigma}+\delta_{\alpha \sigma} \delta_{\beta \gamma}) (2k_1k_2) [2{\text{Im}}{\cal D}_\alpha(k_1^2)]\,[2{\text{Im}}{\cal D}_\beta(k_2^2)].
\eeq
This gives
\begin{align}
\gamma_S
&=4\int \frac{d^3 k}{(2 \pi)^3} [f_N-f_F]\frac{m_{N_1}}{ E_1} \left( \mathcal{C}_\alpha \mathcal{C}_\beta  (m_{N_1}^2)^{\Delta_\alpha+\Delta_\beta-5} \sum_{i \neq 1}\frac{m_{N_1}^2}{m_{N_i}}{\rm Im}[\lambda_{i\alpha} \lambda_{i \beta} \lambda^*_{1 \alpha} \lambda^*_{1 \beta}]\right)\\\no
&=\int \frac{d^3 k}{(2 \pi)^3} [f_N-f_F]\frac{m_{N_1}}{ E_1}\epsilon_1\Gamma_N\\\no
&=\epsilon_1\langle\Gamma_{N_1}\rangle_T(n_N-n_N^{\text{eq}}),
\end{align}
where we used the second line in eq.~\eqref{eq:epsilonasymCFT}. We see that the source rate is $\Gamma_S=\epsilon\langle\Gamma_{N_1}\rangle_T$, as anticipated in \eqref{eq:BoltzeqAsymLV}.

\section{Finite Temperature Corrections}
\label{app:finiteT}

We would like to quantify the impact of finite temperature corrections to the rates. In general this is a very involved task because $T$-dependent effects are strongly model-dependent. Yet, our analysis significantly simplifies if we make a key assumptions: we assume that $n$-point functions remain suppressed compared to the $2$-point functions even at finite $T$. Under this hypothesis all rates can be expressed in terms of $2$-point functions of the CFT even at finite temperature, and our task reduces to estimating the finite $T$-corrections on the latter.

Using the OPE, the finite temperature two point correlation functions relevant to our analysis can be written schematically as \cite{Iliesiu:2018fao}
\beq \label{eq:CFT OPE at finite T}
\langle O(x) O^\dagger (0) \rangle_T \sim \frac{1}{(x^2)^\Delta} \langle 1 \rangle_T +\sum_{O'}\frac{b_{O O^\dagger O'}}{(x^2)^{\Delta-\frac{1}{2}\Delta'}} \langle O' \rangle _T
\eeq
where we suppressed the Lorentz indices for simplicity. The first term, corresponding to the OPE with the identity operator, reproduces the zero temperature result. The reminder encodes the finite temperature corrections, and is proportional to $b_{O O^\dagger O'}$, the OPE coefficients of the operators $O'$. The symbols $\Delta$ and $\Delta'$ correspond to the scaling dimension of $O$ and $O'$, respectively. 

Since $SO(3)$ rotations are preserved by finite temperature and assuming that the SM gauge symmetry remains unbroken at the temperatures of interest, only SM-singlet operators $O'$ with integer spin can contribute, namely those that contain $SO(3)$-singlet components. For those we expect at most $\langle O' \rangle _T \propto T^{\Delta'}$. Therefore, thermal corrections to correlators are controlled by powers of $T x$ and are dominated by the effect of the operators $O'$ of lowest dimension appearing in the OPE. By assumption of the CH models, SM-singlet scalar operators must be marginal or irrelevant, otherwise they would reintroduce the hierarchy problem as their coefficients needs to be tuned to be small. For such operators we thus have $\Delta'\geq4$. The scaling dimensions of spin-1 operators are bounded be $\Delta'\geq3$ by general arguments, whereas a stronger lower bound is found for higher spins. Hence the most relevant corrections are expected to come from vector operators. In the rates evaluated in Appendix \ref{app:nearequilibrium} the relevant distance scale is of order $x\sim1/k\sim 1/m_N$. Hence the temperature corrections are estimated to be suppressed by at least
\beq
\left(\frac{T}{m_N}\right)^3.
\eeq
We conclude that $T$-dependent corrections to the CFT 2-point functions, and ultimately on the rates appearing in the Boltzmann equations, are strongly suppressed as long as we restrict our analysis to low temperatures satisfying $T\ll m_N$.

\section{Intuitive Derivation of the Boltzmann Equations }
\label{app:moreintuitivederivation}

In this appendix, we present an alternative derivation of Boltzmann equations for the high scale genesis. Our discussion in this section may be a bit less rigorous than appendix~\ref{app:nearequilibrium}, but readers may find it more physically intuitive and closer to the standard derivations.

The general form of the Boltzmann equation for the number density of a given quantity $X$ in the early Universe can be written as:
\begin{eqnarray}
    \frac{d n_X}{dt}+3 H n_X = \gamma_{\textrm {prod}, X}-\gamma_{\textrm {dep}, X},
\end{eqnarray}
where $H = \frac{da/dt}{a}$ is the Hubble parameter with $a$ being the scale factor. 
The left-hand side of the equation is the time derivative of the comoving $X$ number density, $\frac{1}{a^3}\frac{d }{d t}(n_X a^3)$. The rates (per unit volume) to produce and deplete $X$ are denoted as  $\gamma_{\textrm {prod}, X}$ and $\gamma_{\textrm {dep}, X}$ respectively.~\footnote{ It may be useful to remind the form of these rates for the particle case (as opposed to CFT). For instance, the production rate takes the form (neglecting Fermi-Dirac v.s.~Bose-Einstein factor $(1 \pm f)$ for final states)
\bea
&& \gamma_{{\rm prod}, X} (ij\cdots \to X a \cdots) = \int \frac{d^3 p_i}{2 E_i (2\pi)^3} \frac{d^3 p_j}{2 E_i (2\pi)^3} \cdots f_i (p_i) f_j (p_j) \cdots \int \int \frac{d^3 p_X}{2 E_X (2\pi)^3} \frac{d^3 p_a}{2 E_a (2\pi)^3} \cdots \nonumber \\
&& \hspace{4cm} (2\pi)^4 \delta^4 (p_i + p_j + \cdots - p_X - p_a - \cdots) \vert A \vert^2
\eea
where $f_i (p_i)$ is the phase space distribution function of the initial state ``$i$''.
}

Our setup involves CFT sector states which may not be described by the conventional particle states. This makes the first principle derivation of rates in general cases non-trivial. For instance, the rate associated with a process ${\rm CFT} \to {\rm particles}$ is challenging to compute when the CFT sector is at finite temperature. 
Instead of tackling this rather technical question, in this work we focus on physically motivated scenarios for which the (approximate form of) relevant rates can be determined.

Before we move forward, however, it maybe worth mentioning existing works in which rates and/or Boltzmann equations involving CFT sector were studied. 
First, zero temperature rates with final state including CFT states can be computed using the formalism of unparticles \cite{Georgi:2007ek, Grinstein:2008qk}. This method can be applied even when $N \to O$ is not an inclusive process. In addition, \emph{thermal averaged rate} with CFT final states can be computed based on general properties of CFT~\cite{Hong:2019nwd,Hong:2022gzo, Chiu:2022bni,Redi:2021ipn}.
Since the derivation of rates with thermal CFT initial states is challenging (even if they can be defined), the Conformal Freeze-in scenario~\cite{Hong:2019nwd, Hong:2022gzo, Chiu:2022bni} focuses on the regime where the backreaction from the CFT sector to external particle sector can be neglected. In such cases, it was shown that Boltzmann equations can be reliably derived \cite{Hong:2019nwd, Hong:2022gzo, Chiu:2022bni}. 

Since the relevant quantities for conformal leptogenesis are the number density of external heavy singlets $N$ ($n_N$)
and lepton asymmetry ($n_{\Delta L}$), we will concentrate on the Boltzmann equations describing the evolution of $n_N$ and $n_{\Delta L}$. 
Furthermore, as we discuss below, for all scenarios considered in this paper, lepton asymmetry is  dominantly generated by the $N$ decay when $N$ is non-relativistic. Therefore, we will restrict our analysis to the regime $T\ll m_N$. 

To leading order in the interaction Hamiltonian $H_{\rm int}=\lambda_{i\alpha}N_i O_{\alpha}+\textrm{h.c.}$, the process to deplete $N$ is $N$ decaying to CFT states associated with the CFT operator $O_\alpha$.
Let us use the notations $N\to O$ and $N\to \bar O$ to denote $N$ decays to CFT states generated by operator $O$ and $\bar O$, respectively.

Then, given $ \gamma_{\textrm {dep}, N}=\gamma(N\to O)+\gamma(N\to \bar O)$ and recalling that $\gamma(N\to O)=\Gamma(N\to O)n_N$ for standard particle decay in the non-relativistic regime,
it is straightforward to get 
\begin{eqnarray}\label{eq:gamma_dep_N}
    \gamma_{\textrm {dep}, N}&=&\gamma(N\to O)+\gamma(N\to \bar O)\nonumber\\
    &=&\Gamma(N\to O)n_N+\Gamma(N\to \bar O)n_N=\Gamma_N\, n_N,
\end{eqnarray}
where we defined $\Gamma_N\equiv\Gamma(N\to O)+\Gamma(N\to \bar O)$.
We note that eq.~\eqref{eq:gamma_dep_N} will in general acquire corrections at finite temperature. However, the correction is suppressed when $N$ is non-relativistic, i.e., the thermal bath is cold enough ($T \ll m_N$).

To obtain the expression for $\gamma_{\textrm {prod}, N}$, we use the equilibrium condition: if $n_N$ is in thermal equilibrium at some temperature $T$, the right-hand side of its Boltzmann equation should vanish. This means $\gamma_{\textrm {prod}, N}= \gamma_{\textrm {dep}, N}$ when $n_N=n_N^{\rm eq}(T)$, where $n_N^{\rm eq}(T)$ is the equilibrium number density for $N$. Since $N\to O$ and $N\to \bar O$ are two independent processes, the standard argument of detailed balance implies that the equality holds for each process. Therefore, from eq.~\eqref{eq:gamma_dep_N}, we find 
\begin{eqnarray}\label{eq:gamma_prod_N}
    \gamma_{{\rm prod},N}&=& \gamma( O\to N)+\gamma( \bar O\to N)\nonumber\\
    &=&\Gamma(N\to O)n_N^{\rm eq}(T)+\Gamma(N\to \bar O)n_N^{\rm eq}(T)=\Gamma_N \, n_N^{\rm eq}(T),
\end{eqnarray}
where $ O\to N$ denotes the process of producing $N$ from CFT states associated with operator $O$.
The above equation expresses that the production of $N$ is mainly from the ``inverse-decay'' of CFT states. Since we consider $T\ll m_N$, the dominant temperature dependence is encoded in $n_N^{\rm eq}(T)$, which has the Boltzmann suppression as expected.

Therefore, the Boltzmann equation for $n_N$ is given as:
\begin{eqnarray}\label{eq:BoltzmanneqN}
    \frac{d n_N}{d t} + 3 H n_N = -\Gamma_N \left( n_N- n^{\rm eq}_N (T) \right),
\end{eqnarray} 
which is basically the same as the one in the standard leptogenesis in weakly coupled theories. It may worth pointing out that a posterior it seems obvious that the form of the equation should be the same as the standard case. This is because the arguments presented here do not rely on the details of the microscopic theory that $N$ couples to, instead it only depends on the assumptions that $N$ is a weakly coupled particles and that the sector it couples to is in thermal equilibrium. 

Now we turn to the Boltzmann equation for $n_L$. Recalling that CFT state associated with the operator $O$ ($\bar O$) carries lepton number $+1 (-1)$, to leading order in the coupling $\lambda$ (i.e.~suppressing higher order terms in $\lambda$) we can understand that the general form of equation is given by 
\beq \label{eq:BE for Delta L}
\frac{d n_L}{d t} + 3 H n_L = \left[ \gamma(N\to O)- \gamma(N\to \bar O) \right] - \left[ \gamma (O \to N) - \gamma (\bar O \to N) \right] + \left[ \Delta L = 2 {\rm \;\; scattering} \right],
\eeq
where the first two terms on RHS correspond to processes of type $N \to {\rm CFT}$, while the second two terms represent processes of the form ${\rm CFT} \to N$. Finally, the last contribution comes from $\Delta L = 2$ ``scattering'' processes, which in our case corresponds to processes of the form $O \leftrightarrow \bar O$. Naively, one may think that such processes come only at higher orders in $\lambda$-expansion. However, we will discuss below that there exists an important subtlety and indeed there is non-trivial contribution even at leading order in $\lambda$ related to ``on-shell subtraction''.

To proceed further, we note that the lepton number asymmetry can change by two effects. The first contribution comes from non-zero \emph{chemical potential} in the lepton number, $\mu$. Physically, this is relevant when the CFT sector carries non-vanishing lepton number asymmetry $n_L$, hence non-zero $\mu/T$, and in that case, even in the absence of CP violation, $n_L$ can evolve through processes ${\rm CFT} \to N$. Specifically, given that CFT sector has more $L=1$ occupation number than $L=-1$ states, this leads to the \emph{washout} of $n_L$. Since $N = \bar N$, $N$-sector does not carry non-trivial $\mu$, hence chemical potential induced effects are irrelevant for $N \to {\rm CFT}$ processes.

The second contribution is from non-vanishing \emph{CP violation}, captured by $\epsilon$. 
With non-vanishing $\epsilon$, the lepton asymmetry can evolve in time, either with or without the chemical potential. In contrast to the chemical potential effects, CPV-induced effects are relevant for both $N \to {\rm CFT}$ and ${\rm CFT} \to N$ processes. 

In this work, we focus on the regime where both $\frac{\mu}{T} \ll 1$ and $\epsilon \ll 1$. This allows us to expand terms on RHS of \eqref{eq:BE for Delta L} in powers of $\frac{\mu}{T}$ and $\epsilon$. The expansion can be applied to the first and last two terms separately, and we work to linear order in each expansion parameter, and suppress terms of order $\mathcal{O} \left( \left( \frac{\mu}{T} \right)^2 \right)$, $\mathcal{O} (\epsilon^2)$ and $\mathcal{O} \left( \epsilon \frac{\mu}{T} \right)$. 
\begin{itemize}
\item[1.] $\left[ \gamma (O \to N) - \gamma (\bar O \to N) \right] \approx \left( \frac{\mu}{T} \right) \gamma_\mu + \epsilon_{\rm prod} \gamma_\epsilon + {\rm higher \text{-} order \text{-} terms}$
    
where $\gamma_\mu$ and $\gamma_\epsilon$ are coefficients of the expansion parameter $\mu/ T$ and $\epsilon$, respectively. Here, the CPV parameter is labelled with a subscript ``prod'' since it corresponds to the CPV parameter associated with $N$-production channel. Each of these ``rate-coefficients'' can be determined as follows. We first discuss $\gamma_\mu$. To this end, we can simply set $\epsilon_{\rm prod}=0$ (recall we keep only the leading order terms) and use the fact that the CFT sector is in thermal equilibrium with temperature $T$ and chemical potential $\mu$ associated with the lepton number. The latter fact implies that the rate $\gamma (O \to N) \propto e^{\mu/T}$ while the rate for the $CP$-conjugate process $\gamma (\bar O \to N) \propto e^{- \mu /T}$. From this, we obtain
\beq
\frac{\gamma (O \to N)}{\gamma (\bar O \to N)}=e^{2 \mu/T}\approx 1+ \frac{2 \mu}{T}.
\eeq

Using this, we get 
\beq
\frac{\mu}{T} \gamma_\mu =  \frac{2\mu}{T} \gamma (\bar O \to N) =  \frac{\mu}{T} \Gamma_N n_N^{\rm eq},
\eeq
where in the last equality, we used \eqref{eq:gamma_dep_N} and \eqref{eq:gamma_prod_N} and discussion in-between.

Next, we discuss $\gamma_\epsilon$. According to the Nanopolous-Weinberg theorem, this term vanishes to leading order in perturbation theory in the lepton-number-violating coupling $\lambda$. At the level of rates, this means that $\epsilon_{\rm prod} \gamma_\epsilon$ vanishes at order $O(|\lambda|^2)$ and the first non-vanishing contribution appears at $O(|\lambda|^4)$. Similarly to the previous discussion, we set $\mu \to 0$. We now want to define the CPV parameter for the $N$ production processes. To this end, we note that two production processes, $O \to N$ and $\bar O \to N$, are $CP$-conjugate to each other, and the CPV parameter $\epsilon_{\rm prod}$ should be proportional to the  difference in the number of $N$ produced from $\bar O \to N$ vs $O \to N$. These considerations suggest that $\epsilon_{\rm prod}$ can be defined as 
\beq
\epsilon_{\rm prod} \equiv \frac{\gamma (\bar O \to N) - \gamma (O \to N)}{\gamma (\bar O \to N) + \gamma (O \to N)} = \frac{\gamma (\bar O \to N) - \gamma (O \to N)}{\gamma_{{\rm prod},N}},
\eeq
and from this we quickly see that $\gamma_\epsilon = \gamma_{{\rm prod},N}$. Notice that our definition is consistent with the conventional definition in which $\epsilon_{\rm prod} > 0$ when net $n_L > 0$ is produced.

\item[2.] $\left[ \gamma (N \to O) - \gamma (N \to \bar O ) \right] \approx \epsilon_{\rm dep} \gamma_{{\rm dep}, N} + {\rm higher \text{-} order \text{-} terms}$ 

There exists no $(\mu / T)$-term because $N$ carries no lepton number. As discussed above, $\gamma_{{\rm dep},N} = \Gamma_N n_N$ and we define the CPV parameter $\epsilon_{\rm dep}$ associated with $N$-depletion processes similarly as before:
\begin{eqnarray}\label{eq:CPV epsilon_dep}
    \epsilon_{{\rm dep}} \equiv \frac{\gamma(N\to O)-\gamma(N\to \bar O)}{\gamma(N\to O)+\gamma(N\to \bar O)}.
\end{eqnarray}
Notice that this reduces to the standard expression in the nonrelativistic regime $m_N \ll T$
\begin{eqnarray}\label{eq:CPV epsilon_dep}
    \epsilon_{{\rm dep}} \approx \frac{\Gamma(N\to O)-\Gamma(N\to \bar O)}{\Gamma(N\to O)+\Gamma(N\to \bar O)}.
\end{eqnarray}

\item[3.] We now discuss $\Delta L = 2$ scattering processes. Specifically, terms showns as ``$\left[ \Delta L = 2 {\rm \;\; scattering} \right]$'' on the RHS of the eq.~\eqref{eq:BE for Delta L} are given by
\beq
\left[ \Delta L = 2 {\rm \;\; scattering} \right] = - 2 \left[ \gamma (O \to \bar O ) - \gamma (\bar O \to O) \right]
\eeq
where the factor 2 is because these correspond to $\Delta L = 2$ processes. To be clear, our notation $O \to \bar O$ means a CFT state created by $O$ converts into another state associated with $\bar O$ via both on-shell and off-shell $N$-exchanges. One crucial point to make is that the on-shell $N$-exchange contributions are already captured by the above discussion, e.g. $\gamma (O \to N)$ and so on. To avoid double counting, therefore, we must subtract these on-shell contributions, and they are leading order pieces. We can express the situation as
\beq
\left[ \Delta L = 2 {\rm \;\; scattering} \right] = - 2 \left[\gamma (O \to \bar O )_{\rm sub} - \gamma (O \to \bar O )_{\rm on \text{-} shell} - \gamma (\bar O \to O )_{\rm sub} + \gamma (\bar O \to O )_{\rm on \text{-} shell} \right]
\eeq
where $\gamma (O \to \bar O )_{\rm sub}$ and $\gamma (\bar O \to O )_{\rm sub}$ correspond to off-shell $N$-exchange parts, hence higher orders in $\lambda$. We will drop these terms from now. The on-shell parts are given by
\bea
&& \gamma (O \to \bar O )_{\rm on \text{-} shell} = \gamma (O \to N) \cdot {\rm Br} (N \to \bar O) = \gamma (O \to N) \; \frac{1}{2} (1 - \epsilon_{\rm dep}) \\
&& \gamma (\bar O \to O )_{\rm on \text{-} shell} = \gamma (\bar O \to N) \cdot {\rm Br} (N \to O) = \gamma (\bar O \to N) \; \frac{1}{2} (1 + \epsilon_{\rm dep}) 
\eea
where we used the definition of $\epsilon_{\rm dep}$. Then, the combination of on-shell pieces can be simplified to
\bea
\left[ \Delta L = 2 {\rm \;\; scattering} \right]_{\rm on-shell} &=& 2 \left[ \gamma (O \to \bar O )_{\rm on \text{-} shell} - \gamma (\bar O \to O )_{\rm on \text{-} shell} \right] \\
&=& - \left[ \epsilon_{\rm prod} \gamma_{{\rm prod}, N} + \epsilon_{\rm dep} \gamma_{{\rm prod}, N}  \right].
\eea

\end{itemize}

Combining all, we arrive at
\beq
\frac{d n_{ L}}{d t} + 3 H n_{L} = - \frac{\mu}{T} \Gamma_N n_N^{\rm eq} +  \epsilon_{\rm dep} \left[ \Gamma_N n_N -  \gamma_{{\rm prod},N}  \right].
\eeq
This equation can be further refined. 
First, we can use $\gamma_{{\rm prod},N} = \Gamma_N n_N^{\rm eq}$ as discussed earlier, and from now on, we will simply denote $\epsilon_{\rm dep}=\epsilon$.
Second, while the precise dependence of $n_{ L}$ on $\mu$ needs more knowledge about the CFT and is model-dependent, generically it is expected that the following linear relation holds for small $\frac{\mu}{T}$: $n_{ L}= \frac{c}{6} \mu T^2$. Here, $c$ is a constant and the overall normalization has been chosen such that a weakly-interacting CFT with only one massless degree of freedom (and its CPT conjugate) carrying unit lepton number  corresponds to $c=1$.

Incorporating these two refinements, we finally arrive at 
\beq \label{eq:BoltzeqAsym}
\frac{d n_{ L}}{d t} + 3 H n_{ L} \approx -\frac{6 }{ c}  \, \frac{n_N^{\rm eq}(T)}{T^3} \Gamma_N \, n_{ L}+ \epsilon \,  \Gamma_N \left( n_N- n^{\rm eq}_N (T) \right).
\eeq

\bibliographystyle{JHEP}
\bibliography{refs}
\end{document}